\documentclass[aps,prd,reprint,11pt,onecolumn,amsmath,amssymb,nofootinbib,groupedaddress,superscriptaddress,floatfix,tightenlines,eqsecnum]{revtex4-2}
\usepackage{macros}

\begin{document}
	\title{Going off the (momentum) grid: A simple implementation of noncold relics in linear Boltzmann solvers}

    \author{Arad Nasiri}
    \email{arad.nasiri@unb.ca}
    \affiliation{Department of Mathematics and Statistics, University of New Brunswick, Fredericton, Canada}
    \affiliation{Blackett Laboratory, Imperial College London, SW7 2AZ, UK}
	\author{Murali M. Saravanan}
    \email{msarav@uw.edu}
    \affiliation{Department of Physics, University of Washington, Seattle, WA, USA}
    \author{Caio B. de S. Nascimento}
    \email{caiobsn@perimeterinstitute.ca}
    \affiliation{Perimeter Institute for Theoretical Physics, Waterloo, Ontario, Canada}
    \affiliation{Department of Physics, University of Washington, Seattle, WA, USA}
\begin{abstract}
Linear Boltzmann solvers typically evolve the perturbations of noncold relics, such as massive neutrinos or hot subcomponents of the dark matter, on a momentum grid, often switching to a fluid approximation on small scales for efficiency. We implement in \class{} the Dispersive Fluid Approximation (DFA), a fluid scheme that captures the scale-dependent sound speed of these species, and reduces errors in the cold dark matter power spectrum by up to an order of magnitude relative to the default \class{} fluid approximation. We also present the first open-source implementation of the Generalized Boltzmann Hierarchy (GBH), which integrates out the momentum dependence of the perturbations of noncold relics altogether. Combined with the DFA and tested on massive neutrinos, the GBH reproduces the matter power spectrum to better than permille accuracy and cosmological parameter constraints to better than $0.05\sigma$, at a computational cost comparable to that of the standard implementation. We argue that the GBH is particularly well suited to relics whose distribution function is expensive to sample on a momentum grid, and illustrate this with a toy model in which a single parameter broadens a thermal distribution. As the broadening increases, the runtime of the standard implementation grows by up to a factor of $\sim 7$, while that of the GBH stays nearly flat, making the GBH up to $\sim 5$ times faster. Our implementation of both the DFA and the GBH in \class{} is publicly available at \href{https://github.com/aradn97/class_GBH}{\faGithub\,\nolinkurl{https://github.com/aradn97/class_GBH}}.
\end{abstract}
	
\maketitle

\bibliographystyle{unsrturl}
	
\section{Introduction}
\label{sec:intro}

We have already learned a lot about the constituents of our Universe and its initial conditions from previous and ongoing cosmological surveys \cite{Planck:2018vyg,AtacamaCosmologyTelescope:2025blo,SPT-3G:2025bzu,eBOSS:2020yzd,DESI:2025zgx,DES:2021wwk,Wright:2025xka,Dalal:2023olq,Li:2023tui}. The future is also very bright, with new instruments promising to deliver unprecedented amounts of high quality data \cite{SimonsObservatory:2018koc,Euclid:2024yrr,LSST:2008ijt,Spergel:2015sza,DESI:2016fyo}, which may well hold the missing key to answer many outstanding open questions in physics, such as the fundamental nature of dark matter \cite{Bertone:2004pz,Cirelli:2024ssz,Baryakhtar:2025jwh} and dark energy \cite{Frieman:2008sn,Weinberg:2013agg}. With such a significant increase in statistical power comes the necessity to further develop both our theoretical models and computational methods, to ensure an efficient unbiased extraction of fundamental physics from the new data.  

Noncold relics, i.e. species whose velocity dispersion remains non-negligible through part of the cosmic history probed by observations, are a highly motivated target of contemporary cosmological surveys \cite{Lesgourgues:2006nd,Boyarsky:2008xj,Green:2019glg,Dvorkin:2019jgs,DePorzio:2020wcz,Xu:2021rwg,Drlica-Wagner:2022lbd,Nascimento:2025bax,Verdiani:2025jcf,Verdiani:2026cpc}. The best-known example is massive neutrinos: at least two of the three known neutrino mass eigenstates must be massive, as required by neutrino (flavor) oscillations, implying a lower bound $M_{\nu} = \sum_{\nu} m_{\nu} \gtrsim 0.06\,\textrm{eV}$ on the sum of the masses \cite{Super-Kamiokande:1998kpq,SNO:2002tuh,Esteban:2024eli,ParticleDataGroup:2024cfk},\footnote{Hereafter we will refer to $M_{\nu}$ as the neutrino mass scale.} while the tightest existing upper bounds, $M_{\nu} \lesssim (0.1-0.2)\,\textrm{eV}$, come from cosmology \cite{Planck:2018vyg,AtacamaCosmologyTelescope:2025blo,SPT-3G:2025bzu} (under the assumption of the concordance $\Lambda$CDM model and its variants).\footnote{We choose to be conservative on the reported upper bound. Recent combinations of Baryon Acoustic Oscillation (BAO) and Cosmic Microwave Background (CMB) data prefer tighter bounds, in some tension with neutrino oscillations, though this is mostly driven by a tension between the BAO and CMB datasets themselves \cite{Loverde:2024nfi,Elbers:2025vlz,Jhaveri:2025neg,Lynch:2025ine}, that may be solved by other means \cite{Jhaveri:2025neg,Sailer:2025lxj,Sullivan:2026tas,Graham:2025dqn,DESI:2025fii,DESI:2025ffm,Reboucas:2024smm,Pompeu:2026zad,Chen:2025mlf,Philcox:2025faf,Lynch:2024hzh,Mirpoorian:2025rfp,Weiner:2026sfm}.} These bounds are not specific to neutrinos: a hot subcomponent of the dark matter that becomes nonrelativistic late enough for its free-streaming scale to remain unresolved by current surveys suppresses the growth of structure in the same way, so that constraints on $M_{\nu}$ can be directly reinterpreted as constraints on such species~\cite{Green:2021xzn}. As observations improve, however, they will become increasingly sensitive to the scale dependence of this suppression~\cite{Boyle:2017lzt,LoVerde:2014pxa,Rogozenski:2023tse}. This is in turn set by the free-streaming scale of each species, which is no longer captured by $M_{\nu}$ alone. Exploiting this information requires accurate, species-specific predictions for the evolution of noncold relic perturbations. Well-motivated candidates in this regime include freeze-in dark matter \cite{Frangipane:2021rtf,Dvorkin:2020xga,Ballesteros:2020adh,DEramo:2020gpr,Decant:2021mhj}, sterile neutrinos \cite{Gelmini:2019wfp,Das:2021pof,Baur:2017stq,Cherry:2017dwu}, and the products of neutrino decays \cite{Chacko:2019nej,FrancoAbellan:2021hdb,Barenboim:2020vrr}, among many others.

The addition of noncold relics is known to significantly complicate the theoretical calculation of cosmological observables of interest, since the momentum dependence of their perturbations must be resolved. For massive neutrinos, state-of-the-art Boltzmann solvers such as the Cosmic Linear Anisotropy Solving System (\texttt{CLASS})~\cite{Lesgourgues:2011re,Blas:2011rf,lesgourgues2011cosmic} and the Code for Anisotropies in the Microwave Background (\texttt{CAMB})~\cite{Lewis:1999bs,Lewis:2026mif} handle this task efficiently at their default precision settings. Producing highly accurate predictions for neutrino transfer functions remains more challenging than for other species, although it can be achieved when adopting very high precision settings, at the cost of an increase in evaluation time by orders of magnitude~\cite{Dakin:2017idt,Lee:2025zym}.\footnote{As expected, the situation is even worse when trying to compute non-linear observables with massive neutrinos. For example, there are a number of practical problems that need to be addressed in order to make the inclusion of massive neutrinos in N-body simulations feasible~\cite{Nascimento:2021wwz,Zennaro:2016nqo,Fidler:2017ebh,Fidler:2018stc,Heuschling:2022rae,Banerjee:2018bxy,Fidler:2018bkg,Elbers:2020lbn,Sullivan:2023ntz,Dakin:2017idt,Euclid:2022qde}, and even then, different approaches produce discrepant results at tens of percent level for the neutrino power spectrum~\cite{Euclid:2022qde,Upadhye:2023bgx,Upadhye:2025glk}.} In practice, since massive neutrinos only make a small fractional contribution to the total energy budget of the Universe after becoming nonrelativistic, one can relax the accuracy requirements on their dynamics and still produce all cosmological observables of interest with errors below the target experimental sensitivity. This substantially reduces runtime, which is key given the tens of thousands of evaluations required in a typical Markov chain Monte Carlo (MCMC) analysis.

In this work we propose a novel implementation of noncold relics in linear Boltzmann solvers, using massive neutrinos as a validation benchmark, since they are the one noncold relic guaranteed to exist and their standard implementation is mature and well tested. Our main motivation is to make parameter exploration of alternative noncold relic species computationally feasible. Our approach is complementary to ongoing efforts to reduce the cost of the Boltzmann solver, and of the sampling procedure as a whole, through GPU acceleration~\cite{Storchi:2025who,Berry:2026dzf} and differentiable solvers that enable gradient-based samplers~\cite{Hahn:2023nvb,Sletmoen:2025fro,Zhou:2026eit}, since these are agnostic to how noncold relics are treated.

Like massive neutrinos, many alternative noncold relic species require resolving the momentum dependence of their perturbations, rather than being described by the simpler moment-truncated hierarchies used for cold dark matter or fully relativistic species. Authors studying such species often modify the \texttt{CLASS ncdm} component directly to obtain their predictions \cite{Garani:2022yzj,DEramo:2020gpr,Decant:2021mhj,FrancoAbellan:2020xnr,FrancoAbellan:2021sxk,FrancoAbellan:2021hdb,Barenboim:2020vrr,Das:2021pof}. The efficiency of \texttt{ncdm} for massive neutrinos, however, is partly tied to the thermal Fermi-Dirac shape of their momentum distribution, for which a tailored Gauss-Laguerre quadrature needs only a handful of momentum points \cite{lesgourgues2011cosmic}. For distorted or non-thermal distributions this advantage is lost, and many more momentum points may be required to reach convergence (see \cref{sec:applications} and \cref{sec:toy_model}). For some of these species, the GBH may offer a simpler and more efficient implementation than the full Boltzmann hierarchy. We discuss the conditions under which this may hold in \cref{sec:generalization}.

Our implementation builds on new promising ideas that have recently emerged in the literature. In \cite{BastosdeSennaNascimento:2021cej} (hereafter \texttt{Paper I}), the Generalized Boltzmann Hierarchy (GBH) was introduced to allow the momentum dependence of massive neutrino perturbations to be effectively integrated out. This offers a simpler alternative to the standard Boltzmann hierarchy: massive neutrinos can be implemented in Boltzmann solvers without having to do any integrals over momentum at the level of perturbations (hence the title of this paper), although these integrals still need to be computed at the background level.\footnote{However, these can be pre-computed and stored in a table, in a way that is agnostic to the neutrino mass and cosmological parameters. See \cref{sec:implementation} for more details.} Additionally, in \cite{Nascimento:2023psl} (hereafter \texttt{Paper II}) a Dispersive Fluid Approximation (DFA) was developed which significantly improves over previous fluid approximation schemes by accounting for the dispersive nature of the neutrino fluid.

In what follows, we briefly summarize our novel implementation of massive neutrino species in the Boltzmann solver \texttt{CLASS}. The neutrino perturbations initially evolve with the GBH, and at some point in the time evolution a switch to the DFA is triggered (a similar strategy is adopted in the standard implementation of massive neutrinos in \texttt{CLASS}). We demonstrate that our implementation incurs only a moderate increase in runtime relative to the standard \texttt{CLASS} \texttt{ncdm} module (in its default precision settings): the time spent computing the evolution of massive neutrino perturbations is $\sim 1.2$ times that of \texttt{ncdm},\footnote{Throughout this paper, our runtime comparisons are based on the massive neutrino overhead $\Delta t = t(M_{\nu})-t(M_{\nu}=0)$, with $t(M_{\nu})$ the total runtime with massive neutrinos included and $t(M_{\nu}=0)$ the runtime in a cosmology without massive neutrinos, while holding fixed all the remaining cosmological parameters. This approximately isolates the time it takes to solve the dynamics of massive neutrino perturbations.} which corresponds to only $\sim 7\%$ in total runtime.

Beyond this, the resulting implementation has a number of attractive features: no integrals over momentum need to be evaluated at the perturbation level (see \cref{sec:implementation}), and the analog of the momentum resolution is controlled per mode by a single truncation parameter, $n_{\textrm{max}}$, rather than by a momentum grid tailored to a given distribution function. This feature may make it better suited to species whose distribution function can be expensive to sample on a momentum grid, whereas the GBH evolves the momentum integrals of the perturbations directly, so its cost need not track the shape of $f_{0}$. We discuss this possibility in \cref{sec:applications} and illustrate it with a toy model in \cref{sec:toy_model}, where the GBH becomes up to $\sim 5$ times faster than \texttt{ncdm} at comparable accuracy as the distribution broadens.

Both the standard Boltzmann hierarchy and the GBH are essentially exact wherever they are directly solved, with truncation errors in the neutrino transfer functions kept $\ll 0.1\%$, so the accuracy of the resulting cosmological observables is instead controlled by the fluid approximation used once truncation becomes impractical. Our dispersive fluid approximation (DFA) is significantly more accurate than the fluid approximations currently implemented in \texttt{CLASS}. Correspondingly, our massive neutrino transfer functions are more accurate than default \texttt{CLASS}, with errors in the neutrino auto power spectrum remaining $\lesssim 10 \%$ at all scales at low redshifts (while default \texttt{CLASS} is off by about a factor of 2 on small scales). Because the DFA is implemented independently of the underlying hierarchy, it can be adopted immediately within the standard \texttt{ncdm} module as well, and since its accuracy gain comes at a modest additional cost of only a few percent in runtime (see \cref{fig:timing_plot}), we advocate for doing so regardless of whether the GBH is also used.

The rest of this manuscript is organized as follows. In \cref{sec:theory} we summarize our theoretical model, based on the combination of GBH$+$DFA. In \cref{sec:implementation} we discuss the more practical aspects related to our numerical implementation of massive neutrinos in the linear Boltzmann solver \texttt{CLASS}. In \cref{sec:testing} we carry out a thorough comparison against standard methods and comment on our results, highlighting the aspects of accuracy and efficiency. In \cref{sec:generalization} we discuss future directions for the GBH, including its potential application to noncold relic species beyond massive neutrinos, illustrated with a toy model of a non-thermal distribution that highlights its attractive features, and an iterative approach to solving the GBH system that is likely to improve its efficiency for any such species. We conclude in \cref{sec:conc}.

\section{Theoretical model}
\label{sec:theory}

Our implementation of massive neutrinos in linear Boltzmann solvers is based on the Generalized Boltzmann Hierarchy (GBH), originally introduced in \texttt{Paper I}. Instead of solving the usual Boltzmann hierarchy for the multipole moments of the distribution function in full phase space \cite{Ma:1995ey}, the GBH allows one to effectively integrate out the momentum dependence of massive neutrino perturbations. The price is a new hierarchy of equations for higher velocity-weighted moments of the distribution function. This yields a two-parameter infinite set of ordinary differential equations with no momentum dependence, which needs to be complemented with a proper truncation scheme. 

For computational efficiency, we switch to a fluid approximation once a mode is well inside the free-streaming horizon, similarly to the approach taken in \texttt{CLASS} \cite{lesgourgues2011cosmic}. However, we switch to an improved Dispersive Fluid Approximation (DFA), first introduced in \texttt{Paper II}. The DFA accounts for the dispersive nature of the neutrino fluid, i.e. the scale dependence of its sound speed, to significantly improve the performance of the fluid approximation in reproducing the exact linear dynamics. We will now describe in more detail these ingredients of our theoretical model. 

\subsection{Generalized Boltzmann Hierarchy (GBH)}
\label{sec:gbh}

\subsubsection{The system of equations}
\label{sec:system_eqs}

We restrict ourselves to linear scalar perturbations around a flat Friedmann-Lemaître-Robertson-Walker (FLRW) metric, in both Newtonian and synchronous gauges, following the notations and conventions of \cite{Ma:1995ey}. The full neutrino phase-space distribution function can be written as $f=f_{0}(q)(1+\Psi)$, where
\begin{equation}
\label{eq:Fermi-Dirac}
    f_{0}(q) = \frac{g_{s}}{(2\pi)^{3}} \frac{1}{1+e^{\frac{q}{T_{0}}}} \,,
\end{equation}
is the standard Fermi-Dirac distribution at the background level. $f_0$ is only a function of the absolute value of the comoving momentum $q$. Also, $g_{s}=2$ counts the number of degrees of freedom (left-handed neutrinos and right-handed antineutrinos), and $T_{0} \approx 1.7 \times 10^{-4} \textrm{eV}$ is the present-day value of the neutrino temperature.\footnote{Neutrino species decouple from the primordial plasma when the Universe is about a second old, hence falling out of thermal equilibrium. Neutrino particles then free-stream along geodesics of the FLRW geometry given by $q=ap=\textrm{const}$, where $p$ is the physical momentum. As a consequence, the relativistic Fermi-Dirac profile of \cref{eq:Fermi-Dirac} is preserved despite the lack of thermalization.} The anisotropies in the neutrino perturbations are captured by $\Psi$, which admits the following multipole decomposition
\begin{equation}
\label{eq:multipoles}
    \Psi(\tau, \vec{k}, \vec{q}) =  \sum_{\ell=0}^{\infty} (-i)^{\ell} (2\ell +1) \Psi_{\ell}(k,q,\tau) P_{\ell}(\hat{k} \cdot \hat{q}) \,,
\end{equation}
where hereafter we work in Fourier space. The starting point of the GBH is to introduce the following set of dynamical variables
\begin{subequations}
\label{eq:gbh_variables}
\begin{align}
    \delta P_{n} \equiv \rho \delta_{n} &= \frac{4\pi}{3} a^{-4} \int_{0}^\infty dq q^{2} f_{0}(q) \epsilon_{q} \left(\frac{q}{\epsilon_{q}}\right)^{2n} \Psi_{0} \,, \label{eq:gbh_variables_delta} \\
    (\rho + P)\theta_{n} &= 4\pi k a^{-4} \int_{0}^\infty dq q^{2} f_{0}(q) \epsilon_{q} \left(\frac{q}{\epsilon_{q}}\right)^{2n+1} \Psi_{1} \,, \label{eq:gbh_variables_theta} \\
    (\rho + P)f_{\ell,n} &= 4\pi \frac{\ell!}{(2\ell -1)!!} a^{-4} \int_{0}^\infty dq q^{2} f_{0}(q) \epsilon_{q} \left(\frac{q}{\epsilon_{q}}\right)^{2n+\ell}\Psi_{\ell} \,. \label{eq:gbh_variables_f}
\end{align}
\end{subequations}
Here, the (comoving) momentum dependence in \cref{eq:multipoles} has been integrated out, and we omit time and wavenumber dependencies for simplicity of notation.\footnote{For example, $\delta P_{n} = \delta P_{n}(\tau, k)$ and $\rho = \rho(\tau)$.} In \cref{eq:gbh_variables}, $\epsilon_{q} = \sqrt{q^{2}+a^{2}m^{2}}$ is the comoving energy of a neutrino particle with mass $m$. Note that $q/\epsilon_{q} = v_{q}$ is the particle velocity. The background energy density is given by
\begin{equation}
\label{eq:back_density}
    \rho = 4\pi a^{-4} \int_{0}^\infty dq q^{2} f_{0}(q) \epsilon_{q} \,,
\end{equation}
and the pressure is
\begin{equation}
\label{eq:back_pressure}
    P = \frac{4\pi}{3} a^{-4} \int dq q^{2} f_{0}(q) \epsilon_{q} \left(\frac{q}{\epsilon_{q}}\right)^{2} \,.
\end{equation}
The equation of state is defined as usual, $\omega = P/\rho$. The standard fluid properties can be obtained from \cref{eq:gbh_variables} by setting $n=0$. In particular, $\delta_{0} = \delta/3$ is a third of the density contrast, $\theta_{0} = \theta$ is the divergence of the velocity field, $f_{2,0} = \sigma$ is the shear stress and $f_{\ell,0}$ are higher multipole moments for $\ell>2$. Setting $n=1$ in \cref{eq:gbh_variables} leads to the pressure perturbation $\delta P_{1} = \delta P$, together with the velocity-weighted analogs of the velocity divergence, $\theta_{1} = \Theta$, and of the shear stress, $\sigma_{1} = \Sigma$, and likewise for the higher multipoles with $\ell > 2$. The $n \geq 2$ case produces even higher velocity weights of the same multipole moments. It will be convenient to set $\theta_{n} = kf_{1,n}$ and $\sigma_{n} = f_{2,n}$ in the following. 

This particular choice of dynamical variables given by \cref{eq:gbh_variables} is justified by the fact that one can write a closed system of equations involving only these quantities. It can be derived by taking derivatives of \cref{eq:gbh_variables} with respect to conformal time (which we denote with a prime, e.g. $\rho'$ is the derivative of the energy density $\rho$).\footnote{In the process one needs to use the dynamical equations for $\Psi_{\ell}$. We refer the reader to \cite{Ma:1995ey}.} In the Newtonian gauge this reads
\begin{subequations}
\label{eq:gbh_new}
\begin{align}
    \delta'_{n} &= -(2n-3\omega) \mathcal{H} \delta_{n} + (2n-1) \mathcal{H} \delta_{n+1} -\frac{1}{3}(1+\omega)\theta_{n} \nonumber \\
    &\quad + \left[(2n+3)\omega_{n} - (2n-1)\omega_{n+1}\right]\phi' \,, \label{eq:gbh_new_delta} \\
    \theta'_{n} &= - \left[(2n+1-3\omega)\mathcal{H} + \frac{\omega'}{1+\omega}\right]\theta_{n} + 2n \mathcal{H} \theta_{n+1} + \frac{1}{1+\omega} k^{2} \delta_{n+1} - k^2 \sigma_{n} \nonumber \\
    &\quad + \frac{1}{1+\omega} \left[(2n+3) \omega_{n} -(2n-1)\omega_{n+1}\right]k^2 \psi \,, \label{eq:gbh_new_theta} \\
    f'_{\ell,n} &= - \left[(2n+\ell-3\omega)\mathcal{H} + \frac{\omega'}{1+\omega}\right] f_{\ell,n} + (2n+\ell-1) \mathcal{H} f_{\ell,n+1} \nonumber \\
    &\quad + \frac{\ell^2}{4\ell^2-1}kf_{\ell-1,n+1} - kf_{\ell+1, n} \,, \label{eq:gbh_new_f}
\end{align}
\end{subequations}
where $\phi$ and $\psi$ are the metric fluctuations in the Newtonian gauge, $\mathcal{H} = a'/a$ is the conformal Hubble rate and
\begin{equation}
\label{eq:eos}
    P_{n} \equiv \rho \omega_{n} = \frac{4\pi}{3} a^{-4} \int dq q^{2} f_{0}(q) \epsilon_{q} \left(\frac{q}{\epsilon_{q}} \right)^{2n} \,,
\end{equation}
defines the higher velocity-weighted background pressures $P_{n}$ (or equations of state $\omega_{n}$), with $\omega_{0}= 1/3$ and $\omega_{1}= \omega$. The synchronous gauge version of \cref{eq:gbh_new} is 
\begin{subequations}
\label{eq:gbh_sync}
\begin{align}
    \delta'_{n} &= -(2n-3\omega) \mathcal{H} \delta_{n} + (2n-1) \mathcal{H} \delta_{n+1} -\frac{1}{3}(1+\omega)\theta_{n} \nonumber \\
    &\quad - \frac{1}{6} \left[(2n+3)\omega_{n} - (2n-1)\omega_{n+1}\right]h' \,, \label{eq:gbh_sync_delta} \\
    \theta'_{n} &= - \left[(2n+1-3\omega)\mathcal{H} + \frac{\omega'}{1+\omega}\right]\theta_{n} + 2n \mathcal{H} \theta_{n+1} + \frac{1}{1+\omega} k^{2} \delta_{n+1} - k^2 \sigma_{n} \,, \label{eq:gbh_sync_theta} \\
    f'_{\ell,n} &= - \left[(2n+\ell-3\omega)\mathcal{H} + \frac{\omega'}{1+\omega}\right] f_{\ell,n} + (2n+\ell-1) \mathcal{H} f_{\ell,n+1} \nonumber \\
    &\quad + \frac{\ell^2}{4\ell^2-1}kf_{\ell-1,n+1} - kf_{\ell+1, n} + \frac{\delta^{K}_{\ell, 2}}{1+\omega} \left[(2n+5)\omega_{n+1} - (2n+1)\omega_{n+2}\right]\left(\frac{2h'}{15}+\frac{4\eta'}{5}\right)\,, \label{eq:gbh_sync_f}
\end{align}
\end{subequations}
with $h$ and $\eta$ the synchronous gauge metric perturbations, and $\delta^{K}_{\ell,2}$ the Kronecker delta (which is unity for $\ell=2$, and zero otherwise). The system of Eqs.~\eqref{eq:gbh_new} (or Eqs.~\eqref{eq:gbh_sync} in the synchronous gauge) constitutes the Generalized Boltzmann Hierarchy (GBH).\footnote{A similar expansion in powers of the velocity is used in \texttt{CAMB} as an approximation to the neutrino dynamics after it transitions into the non-relativistic regime \cite{Lewis:2002nc}. We also note that there should be a mapping between the GBH and the linear Vlasov Perturbation Theory (VPT) \cite{Garny:2022tlk}. These two approaches differ in that the GBH organizes the expansion in terms of the moments of the distribution function, while VPT is based on the cumulant expansion.} The key advantage of the GBH over the standard Boltzmann hierarchy is that the momentum dependence of the neutrino perturbations has been integrated out completely. This means that the GBH implementation of massive neutrinos (and other noncold relics) is much simpler than the standard approach.

There are two more ingredients we need to specify to enable a numerical implementation of the GBH. First, we need to set up the initial conditions for the dynamical variables in \cref{eq:gbh_variables}. Second, the system of \cref{eq:gbh_new,eq:gbh_sync} is of infinite dimensionality, so we need to provide a suitable truncation scheme. We next address both of these points.  

\subsubsection{Initial conditions}
\label{sec:ics}

In the relativistic regime, the higher velocity-weighted quantities are completely redundant, in the sense that $\delta_{n} = \delta_{0}$, $\theta_{n} = \theta_{0}$ and $f_{\ell, n} = f_{\ell, 0}$ for all values of n. In that limit, all neutrino particles travel at the speed of light, and we simply recover the usual hierarchy of equations for radiation. It is then straightforward to set the initial conditions for the GBH, as long as the integration starts deep within the relativistic regime. This assumption was made in \texttt{Paper I}, but in the present work we remain agnostic about when the integration starts in comparison to the transition between the relativistic and non-relativistic regimes. 

On super-horizon scales, the multipole moments of the neutrino distribution function become
\begin{subequations}
\label{eq:super_horizon}
\begin{align}
    \Psi_{0} &= - \frac{1}{4} \delta^{\textrm{ur}} \frac{d\log f_{0}}{d\log q} \,, \label{eq:super_horizon_0} \\
    \Psi_{1} &= - \frac{\epsilon_{q}}{3qk} \theta^{\textrm{ur}} \frac{d\log f_{0}}{d\log q} \,, \label{eq:super_horizon_1} \\
    \Psi_{2} &= -\frac{1}{2} \sigma^{\textrm{ur}} \frac{d\log f_{0}}{d\log q} \,, \label{eq:super_horizon_2} \\
    \Psi_{\ell} &\approx 0 \ \forall \ \ell>2 \,, \label{eq:super_horizon_higher}
\end{align}
\end{subequations}
where the superscript ur, e.g. in $\delta^{\textrm{ur}}$, stands for the corresponding quantity for a strictly ultra-relativistic neutrino species. We may now substitute \cref{eq:super_horizon} into \cref{eq:gbh_variables} to obtain, after integration by parts and a few straightforward algebraic manipulations
\begin{subequations}
\label{eq:gbh_ics}
\begin{align}
    \delta_{n} &= \frac{1}{4} \left[(2n+3)\omega_{n}-(2n-1)\omega_{n+1}\right]\delta^{\textrm{ur}} \,, \label{eq:gbh_ics_delta} \\
    \theta_{n} &= \frac{1}{1+\omega} \left[(2n+3)\omega_{n}-(2n-1)\omega_{n+1}\right] \theta^{\textrm{ur}} \,, \label{eq:gbh_ics_theta} \\
    \sigma_{n} &= \frac{1}{1+\omega} \left[(2n+5)\omega_{n+1}-(2n+1)\omega_{n+2} \right] \sigma^{\textrm{ur}} \,, \label{eq:gbh_ics_sigma} \\
    f_{\ell, n} &\approx 0 \ \forall \ \ell > 2 \,. \label{eq:gbh_ics_higher}
\end{align}
\end{subequations}
We adopt Eqs.~\eqref{eq:gbh_ics} as initial conditions for the GBH in terms of the known initial conditions for an ultra-relativistic neutrino species. Note that this is agnostic to the gauge choice and hence works in both Newtonian and synchronous gauges.   
\subsubsection{Truncation scheme}
\label{sec:truncation}

We now specify how the GBH is truncated into a finite system of ordinary differential equations. This first involves a choice of $\ell_{\textrm{max}}$ and $n_{\textrm{max}}$ which dictates when the truncation happens. In other words, we only include the dynamical equation for $f_{\ell, n}$ when $\ell \leq \ell_{\textrm{max}}$ and $n \leq n_{\textrm{max}}$. The dimension of the resulting system of equations is $(\ell_{\textrm{max}}+1) \times (n_{\textrm{max}}+1)$. The optimal choice is given by the smallest system of equations that still yields converged neutrino transfer functions for a given neutrino mass $m$, comoving scale $k$, and target accuracy. The natural time variable associated to the massive neutrino dynamics is (see Appendix A in \texttt{Paper I})
\begin{equation}
\label{eq:time_variable}
    y = k\Tau = k \int_{\tau_{\textrm{i}}} d\tau \frac{q}{\epsilon_{q}} \,,
\end{equation}
where $\Tau$ is the neutrino horizon, i.e. the integral of the particle velocity over conformal time $\tau$, starting from some initial value $\tau_{\textrm{i}}$. We evaluate this at $q/T_{0} = 3$, close to the mean momentum of the Fermi-Dirac distribution $q^{2}f_{0}(q)$ from \cref{eq:Fermi-Dirac}, and this choice can be adapted to other distribution functions in the code (see \cref{sec:implementation}). We find that the following choices
\begin{subequations}
\label{eq:trunc_size}
\begin{align}
    \ell_{\textrm{max}} &= \frac{y}{2} + 2 \,, \label{eq:trunc_size_lmax} \\
    n_{\textrm{max}} &= \frac{y^{1.6}}{5} + 3 \,, \label{eq:trunc_size_nmax}
\end{align}
\end{subequations}
with the right-hand sides rounded up to the nearest integer, yield both the neutrino density and velocity transfer functions with permille-level accuracy up to a time $y$.

The second step in defining the truncation scheme is to provide approximations for the higher order quantities in terms of lower order ones. The truncation with respect to $\ell_{\textrm{max}}$ is straightforward and reads, following \texttt{Paper I}
\begin{equation}
\label{eq:trunc_lmax}
    f_{\ell_{\textrm{max}}+1,n} \approx (\ell_{\textrm{max}}+1)\left(\frac{1}{k\tau} f_{\ell_{\textrm{max}},n} - \frac{\ell_{\textrm{max}}}{4\ell_{\textrm{max}}^{2}-1} f_{\ell_{\textrm{max}}-1,n+1} \right) \,.
\end{equation}
Since $f_{\ell,n}$ is gauge invariant for $\ell \geq 2$, \cref{eq:trunc_lmax} is also gauge invariant provided that $\ell_{\textrm{max}} \geq 3$, and it can hence be applied in both Newtonian and synchronous gauges.\footnote{We will only consider the case of $\ell_{\textrm{max}} \geq 3$ when solving the GBH.} The truncation with respect to $n_{\textrm{max}}$ is given by
\begin{subequations}
\label{eq:trunc_nmax}
\begin{align}
    \frac{\delta_{n_{\textrm{max}}+1}}{\delta_{n_{\textrm{max}}}} &\approx \Lambda_{0} \equiv \frac{\left[(2n_{\textrm{max}}+5)\omega_{n_{\textrm{max}}+1} - (2n_{\textrm{max}}+1)\omega_{n_{\textrm{max}}+2} \right]}{\left[(2n_{\textrm{max}}+3)\omega_{n_{\textrm{max}}} - (2n_{\textrm{max}}-1)\omega_{n_{\textrm{max}}+1} \right]} \,, \label{eq:trunc_nmax_delta} \\
    \frac{f_{\ell, n_{\textrm{max}}+1}}{f_{\ell, n_{\textrm{max}}}} &\approx \Lambda_{\ell} \equiv \frac{\left[(2n_{\textrm{max}} + 2\ell +3)\omega_{n_{\textrm{max}}+\ell} - (2n_{\textrm{max}}+2\ell -1)\omega_{n_{\textrm{max}}+\ell + 1} \right]}{\left[(2n_{\textrm{max}}+2\ell +1)\omega_{n_{\textrm{max}}+\ell -1} - (2n_{\textrm{max}}+2\ell -3)\omega_{n_{\textrm{max}}+\ell} \right]} \ \forall \ell \geq 1 \,, \label{eq:trunc_nmax_f}
\end{align}
\end{subequations}
following \texttt{Paper I} (note that $\Lambda_{1}=\Lambda_{0}$). It can be shown that the truncation in \cref{eq:trunc_nmax} is also gauge invariant,\footnote{\label{fn:trunc_nmax} The reason for this is as follows. The approximations in \cref{eq:trunc_nmax} derive from the assumption that the momentum dependence of the multipole moments of the perturbed distribution function is given by $\Psi_{0} \propto d\log f_{0}/d\log q$ and $\Psi_{\ell} \propto (q/\epsilon_{q})^{\ell-2} d\log f_{0}/d\log q$ (see Appendix A in \texttt{Paper I} for more details). The gauge invariance of the truncation scheme then follows as a consequence of the assumed momentum dependence being preserved by a gauge transformation.} so that it can be applied in both Newtonian and synchronous gauges. Note, however, that this momentum dependence (derived in Appendix A of \texttt{Paper I}) is only accurate on large scales, where the perturbations are driven by gravity rather than by free streaming. As $k\Tau$ grows, free streaming reshapes the momentum dependence, so that the error introduced at $n_{\textrm{max}}$ is no longer negligible and propagates down the hierarchy, accumulating over the integration. This is why $n_{\textrm{max}}$ must grow so rapidly with $y$ in \cref{eq:trunc_size}, and it is a likely origin of the numerical instability of the GBH deep inside the horizon, which we discuss in \cref{app:trigger}. 

We may now combine the system of Eqs.~\eqref{eq:gbh_new} [or Eqs.~\eqref{eq:gbh_sync}], with the truncation schemes of \cref{eq:trunc_lmax,eq:trunc_nmax}, to numerically solve for the GBH, with initial conditions provided by \cref{eq:gbh_ics}. This provides a simple and accurate approach to evolve linear massive neutrino fluctuations, as first demonstrated in \texttt{Paper I}, at both large and intermediate comoving scales (when compared to the neutrino horizon). On the other hand, the GBH becomes prohibitively expensive on small scales since $n_{\textrm{max}} \propto y^{1.6} \propto k^{1.6}$ from \cref{eq:time_variable,eq:trunc_size_nmax} (and may not even converge, see \cref{app:trigger}). Our strategy will then be to solve the GBH only up to some finite time $y$, and then switch to a Dispersive Fluid Approximation (DFA) which we now describe in more detail.

\subsection{Dispersive Fluid Approximation (DFA)}
\label{sec:fa}

The DFA was first introduced in \texttt{Paper II}, where it was shown that the standard fluid approximation for massive neutrino species can be significantly improved by taking its dispersive nature into account. The basic idea is to truncate the GBH at $\ell_{\textrm{max}}=2$ and $n_{\textrm{max}}=0$ at sufficiently small scales. The motivation for doing so is threefold. First, at scales below the neutrino horizon, neutrino anisotropies are washed out by free-streaming and hence one can relax the accuracy requirements in the evaluation of the neutrino perturbations, in favor of a more efficient approximate scheme. Second, after the transition to the nonrelativistic regime neutrinos become progressively colder, naturally suppressing higher-order terms in both the $\ell$ and $n$ expansions. Third, on small scales it becomes prohibitively expensive to solve the full GBH as we investigate in \cref{app:trigger}.

In the case of $\ell_{\textrm{max}}=2$ and $n_{\textrm{max}}=0$, \cref{eq:gbh_new} simplifies to
\begin{subequations}
\label{eq:FA_new}
\begin{align}
    \delta' &= -(1+\omega) (\theta-3\phi') - 3\mathcal{H} (c_{s}^2-\omega)\delta \,, \label{eq:FA_new_delta} \\
    \theta' &= - \left[(1-3\omega)\mathcal{H} +\frac{\omega'}{1+\omega} \right] \theta + \frac{c_{s}^{2}}{1+\omega} k^2 \delta - k^{2}\sigma + k^2 \psi \,, \label{eq:FA_new_theta} \\
    \sigma' &= -\left[(2-3\omega)\mathcal{H} +\frac{\omega'}{1+\omega} \right]\sigma + \mathcal{H} \Sigma +\frac{4}{15} \Theta -kf_{3} \,, \label{eq:FA_new_sigma}
\end{align}
\end{subequations}
in the Newtonian gauge. Similarly, in the synchronous gauge, we have
\begin{subequations}
\label{eq:FA_sync}
\begin{align}
    \delta' &= -(1+\omega) \left(\theta + \frac{h'}{2}\right) - 3\mathcal{H} (c_{s}^2-\omega)\delta \,, \label{eq:FA_sync_delta} \\
    \theta' &= - \left[(1-3\omega)\mathcal{H} +\frac{\omega'}{1+\omega} \right]\theta + \frac{c_{s}^{2}}{1+\omega} k^2 \delta - k^{2}\sigma \,, \label{eq:FA_sync_theta} \\
    \sigma' &= -\left[(2-3\omega)\mathcal{H} +\frac{\omega'}{1+\omega} \right]\sigma + \mathcal{H} \Sigma +\frac{4}{15} \Theta -kf_{3} + \frac{1}{1+\omega} \left(\omega_{1} -\frac{\omega_{2}}{5}\right)\left(\frac{2h'}{3} + 4\eta'\right)\,. \label{eq:FA_sync_sigma}
\end{align}
\end{subequations}
In \cref{eq:FA_new,eq:FA_sync}, $c_{s}^{2} = \delta P/\delta \rho$ stands for the effective sound speed in the appropriate gauge. These systems of equations are not closed as they are, since there are no dynamical equations for the quantities $c_{s}^{2}$, $\Sigma$, $\Theta$, and $f_{3}$ appearing on their right-hand side. The basic strategy underlying the fluid approximation is to provide approximate formulas for these quantities in terms of the density, velocity, and shear, hence turning \cref{eq:FA_new,eq:FA_sync} into a closed system. 

To estimate the sound speed we follow the approach of \texttt{Paper II}, which is based on an analytic calculation of its large and small scale limits (when compared to the free-streaming scale). In the Newtonian gauge, it reads
\begin{equation}
\label{eq:cs_fa_trunc}
    c_{s}^{2}(a,k) = c_{\textrm{g}}^{2}(a) + e^{-\frac{4}{3} \frac{k_{\textrm{fs}}(a)}{k}} \left[c_{\textrm{asp}}^{2}(a) - c_{\textrm{g}}^{2}(a)\right] \,.
\end{equation}
Here, $c_{\textrm{g}}^{2}(a)$ and $c_{\textrm{asp}}^{2}(a)$ are the adiabatic and asymptotic sound speeds which correspond to the large and small scale limits of $c_{s}^{2}(a,k)$, respectively,\footnote{This is, in fact, a gauge invariant statement. The exact transition $c_{\textrm{g}}^{2}(a) \to c_{\textrm{asp}}^{2}(a)$ as $k/k_{\textrm{fs}}$ goes from zero to infinity depends on the gauge choice, but the strict limits at $k/k_{\textrm{fs}} \to 0, \infty$ do not.} and $k_{\textrm{fs}}(a)$ is the neutrino free-streaming scale. We refer the reader to \texttt{Paper II} for the exact definition of these quantities. It is straightforward to map \cref{eq:cs_fa_trunc} into the synchronous gauge, where it reads
\begin{equation}
\label{eq:cs_fa_trunc_sync}
\frac{\delta P}{\rho} = c_{s}^{2} \delta =  c_{s,\textrm{New}}^{2} \delta + \alpha(c_{s,\textrm{New}}^{2}-c_{\textrm{g}}^{2}) \frac{\rho'}{\rho} \,,
\end{equation}
with $c_{s,\textrm{New}}^{2}$ the Newtonian gauge approximation for the effective sound speed, written in \cref{eq:cs_fa_trunc}. In \cref{eq:cs_fa_trunc_sync}, $\alpha = (h'+6\eta')/2k^{2}$ when mapping to the synchronous gauge, while in general it corresponds to the conformal time shift of an arbitrary gauge transformation (starting from the Newtonian gauge). 

Next, \texttt{Paper II} introduces a scale-dependent phenomenological parametrization of the shear stress that is tuned to produce accurate results for the fluid approximation, while only solving for the density and velocity fields. This strategy is particularly effective in capturing the growth of massive neutrino perturbations that follows after the transition to the nonrelativistic regime. However, in the present set-up where the DFA is activated after a switch from solving the full Boltzmann equation, we find that this phenomenological parametrization induces discontinuities in the shear stress at the switch that can lead to instabilities in the numerical integration. To avoid this, we instead decided to follow previous approaches \cite{lesgourgues2011cosmic, Hu:1998kj}, which keep the shear stress as a dynamical variable and hence do not suffer from the same discontinuities. This latter approach also more accurately captures the dynamics of massive neutrino perturbations while in the relativistic regime, at the cost of performing slightly worse after the transition to nonrelativistic.

With this in mind, we still need to provide approximate expressions for $\Sigma$, $\Theta$, and $f_{3}$ in order to close the system of equations. We simply adopt the choices made in Ref.~\cite{lesgourgues2011cosmic}, which we review in the next few paragraphs for the reader's convenience. We emphasize that the overall accuracy of the fluid approximation is mostly driven by the choice of the sound speed, and is not very sensitive to the modeling of the remaining quantities. Setting $\ell_{\textrm{max}}=2$ and $n=0$ in \cref{eq:trunc_lmax} yields
\begin{equation}
\label{eq:trunc_lmax_fa}
    f_{3} \approx \frac{3}{k\tau} \sigma - \frac{2}{5} \frac{\Theta}{k} \,,
\end{equation}
in the Newtonian gauge. The corresponding synchronous gauge formula can be derived by applying a gauge transformation, and reads
\begin{equation}
\label{eq:trunc_lmax_fa_sync}
    f_{3} \approx \frac{3}{k\tau} \sigma - \frac{2}{5k} (\Theta + 3c_{\textrm{g}}^{2}\alpha k^{2}) \,.
\end{equation}
For the higher velocity-weight velocity divergence, we take
\begin{equation}
\label{eq:trunc_lmax_fa_theta}
    \Theta \approx 12 \frac{\omega}{1+\omega} c_{\textrm{g}}^2 \theta  \,,
\end{equation}
in the Newtonian gauge and 
\begin{equation}
\label{eq:trunc_lmax_fa_theta_sync}
    (\Theta + 3c_{\textrm{g}}^{2}\alpha k^{2}) \approx 12 \frac{\omega}{1+\omega} c_{\textrm{g}}^2 (\theta + \alpha k^2) \,,
\end{equation}
in the synchronous gauge.\footnote{This is equivalent to the viscosity parameter $c_{\textrm{vis}}^{2} = 3\omega c_{\textrm{g}}^{2}$ of the \texttt{CLASS} fluid approximation \cite{lesgourgues2011cosmic}.} Finally, for the velocity-weighted shear, we assume 
\begin{equation}
\label{eq:trunc_fa_sigma}
    \Sigma \approx \frac{\omega_{2}}{\omega_{1}} \sigma \,,
\end{equation}
which is a gauge invariant formula and hence applies in both Newtonian and synchronous gauges. With these closures, the metric enters the shear equation only through the combination $\theta + \alpha k^{2}$ in the synchronous gauge, and not at all in the Newtonian gauge. In our implementation we instead follow the recipe of Ref.~\cite{lesgourgues2011cosmic} adopted in \texttt{CLASS}, which replaces $\theta + \alpha k^{2}$ by $\theta + h'/2$ in the synchronous gauge, i.e.\ drops the $6\eta'$ term, since this slightly improves the matter power spectrum. In the Newtonian gauge, $\theta$ is correspondingly replaced by $\theta - 6\phi'$. We are now ready to discuss in more detail our new numerical implementation of massive neutrinos in the linear Boltzmann solver \texttt{CLASS}. 

\section{Implementation in \texttt{CLASS}}
\label{sec:implementation}

Here for the first time we provide an open source implementation of the GBH inside the Boltzmann code \texttt{CLASS} \cite{Blas:2011rf,lesgourgues2011cosmic}. So far by default, \texttt{CLASS} uses a component called \texttt{ncdm} for massive neutrinos. At the background level, \texttt{ncdm} integrates the Fermi-Dirac distribution over momentum space using a quadrature strategy. This is done at every redshift $z$ for finding the energy density $\rho_{\texttt{ncdm}}(z)$, the pressure $P_{\texttt{ncdm}}(z)$, and the so-called pseudo-pressure $\mathfrak{p}(z)  $ \cite{lesgourgues2011cosmic} (which equals $P_2$ in the GBH at the background level). At the perturbations level, \texttt{ncdm} uses a binning of the momentum space to evolve $\Psi_\ell(q)$ as a function of the conformal time. Once again, momentum integrals have to be carried out in two cases: either when there is a transition from the Boltzmann hierarchy to a fluid approximation, or more importantly, when total stress-energy tensor quantities like $\delta \rho_{\rm total}$, $\delta P_{\rm{total}}$, $(\rho+P)\theta_{\rm total}$, and $(\rho+P)\sigma_{\rm total}$ are needed. The total stress energy tensor in turn is used in \texttt{CLASS}  to compute the metric perturbations through Einstein equations.

As a result, \texttt{ncdm} needs to perform momentum integrals at each conformal time step. Each one of these integrals is performed efficiently using a quadrature sum. Nevertheless, their repeated evaluation adds bookkeeping and requires a momentum grid that must be sampled accurately, although at a modest computational cost. If only there was a way to integrate out the momentum dependence of massive neutrino perturbations, one would avoid the momentum grid and its integrals altogether. This is exactly what the GBH accomplishes: it consists of a system of coupled ODEs labeled by $n$, $\ell$, and $k$, well suited for implementation in a Boltzmann code. Energy-momentum tensor quantities like the energy density, pressure, velocity and shear can be readily read off as the lowest order moments in $n$ and $\ell$ with no need for any additional integration.

To implement the GBH at the background level, one option is to use the following evolution equation, which can be derived by taking the derivative of \cref{eq:eos} with respect to conformal time,
\begin{equation}
    P_n'=-(2n+3)\mathcal{H}P_n+(2n-1)\mathcal{H}P_{n+1},
\end{equation}
and truncate the background hierarchy at some $n_{\textrm{max}}$. This is indeed a viable approach for non-cold dark matter models where the species remains non-relativistic at all times (see Ref.~\cite{Albertini:2024yfm} for an example). In these cases, $P_{n_{\textrm{max}}+1}$ can be safely set to zero. However, this does not work for massive neutrinos because they are relativistic at early times, and so $\omega_n\approx1/3$ for all $n$. The higher moments are not suppressed, and so a non-trivial truncation scheme has to be used for $P_{n_{\textrm{max}}+1}$. 

For massive neutrinos, there are two solutions to bypass this problem. The first is to employ the same quadrature strategy as \texttt{ncdm} to compute $P_n$ up to $n_{\textrm{max}}$, and store the corresponding $\omega_n$ to be used in perturbation equations. This option is used if the user passes the input parameter \colorbox{lightgray}{\texttt{gbh\_use\_table = 0}}. The second method is to compute, once for all, the integrals that give $\hat{\rho}= \rho/T^{4}$ and $\omega_n$ as functions of $x=m/T$, with $T=T_{0}/a$ the neutrino temperature in terms of its present-day value $T_{0}$, and store them in tables. This can be done with arbitrarily high precision. Then in the code, we first interpolate the precomputed tables for all but very large values of $x$. For values of $x$ beyond $x_{\textrm{max}}$ of the table, we are deep in non-relativistic regime and we can use simple power-law extrapolations: $\hat{\rho}\propto x$ and $\omega_n\propto x^{-2n}$. All GBH background quantities can then be read for any choice of neutrino mass and redshift. This approach is both faster and more accurate, and is therefore set as the default choice of the input parameter \colorbox{lightgray}{\texttt{gbh\_use\_table = 1}}.

The code also accepts distribution functions other than \cref{eq:Fermi-Dirac}. Mirroring the corresponding \ncdm{} options, setting \colorbox{lightgray}{\texttt{use\_gbh\_psd\_files = 1}} reads $f_{0}(q)$ from a two-column table, given by \colorbox{lightgray}{\texttt{gbh\_psd\_filenames}}, with the same normalization as the \ncdm{} tables. Since the precomputed tables of $\omega_n$ described above assume a Fermi-Dirac distribution, the code then sets \texttt{gbh\_use\_table}~$=0$ and computes all background quantities by quadrature. The perturbation equations themselves are unchanged, since $f_{0}$ enters them, their initial conditions, and their truncation only through $\omega_n$. A second parameter, \colorbox{lightgray}{\texttt{gbh\_q\_horizon}}, sets the momentum $q/T_{0}$ at which the free-streaming horizon $\Tau$ of \cref{eq:time_variable} is evaluated, and hence the truncation scheme of \cref{eq:trunc_size} and the time of the switch to the DFA. Its default value of $3$ is the choice made in \cref{sec:truncation}. Setting \colorbox{lightgray}{\texttt{gbh\_q\_horizon = auto}} instead uses the mean momentum of the given distribution, $\langle q \rangle = \int dq\, q^{3} f_{0} / \int dq\, q^{2} f_{0}$, which equals $3.15\,T_{0}$ for Fermi-Dirac.

At the level of  perturbations, we simply provide \texttt{CLASS} with the GBH equations, in both Newtonian and synchronous gauges, to be evolved alongside other perturbation equations. These are evolved to solve for $\delta_n$ and $f_{\ell,n}$ (with the understanding that $\theta_n=kf_{\ell=1,n}$) as functions of conformal time. For each $k$, the value of $y=k\Tau$ increases with time. So, according to \cref{eq:trunc_size}, ever larger $n_{\textrm{max}}$ and $\ell_{\textrm{max}}$ would be needed to ensure precision. To prevent this from slowing down the GBH, we switch to the Dispersive Fluid Approximation (DFA) when $y$ crosses the parameter \colorbox{lightgray}{\texttt{gbh\_FA\_trigger}} in the code, for which we have selected the default value \colorbox{lightgray}{\texttt{gbh\_FA\_trigger}=15}. Before the switch, we use the truncation scheme outlined in \cref{sec:truncation}. Note that we are using a fixed $n_{\textrm{max}}$ and $\ell_{\textrm{max}}$ for a given $k$ at \textit{all times} $y$, $0\leq y\leq \min[y(\tau_0),\texttt{gbh\_FA\_trigger}]$. We call this the \textit{aggressive truncation scheme}, because it uses a smaller $n_{\textrm{max}}$ and $\ell_{\textrm{max}}$ at larger scales where $y(\tau_0)<$ \texttt{gbh\_FA\_trigger}:
\begin{subequations}
\label{eqn:nmax_lmax_vs_y}
\begin{align}
    \ell_{\textrm{max}} &= \frac{y(\tau_0)}{2} + 2 \,, \label{eqn:nmax_lmax_vs_y_lmax} \\
    n_{\textrm{max}} &= \frac{y(\tau_0)^{1.6}}{5} + 3 \,. \label{eqn:nmax_lmax_vs_y_nmax}
\end{align}
\end{subequations}
At smaller scales, \texttt{gbh\_FA\_trigger} $<y(\tau_0)$, and so
\begin{subequations}
\label{eqn:conservative_trunc}
\begin{align}
    \ell_{\textrm{max}} &= \frac{\texttt{gbh\_FA\_trigger}}{2} + 2 \,, \label{eqn:conservative_trunc_lmax} \\
    n_{\textrm{max}} &= \frac{\texttt{gbh\_FA\_trigger} ^{1.6}}{5} + 3 \,. \label{eqn:conservative_trunc_nmax}
\end{align}
\end{subequations}
In principle, one may even choose to adjust $n_{\textrm{max}}$ and $\ell_{\textrm{max}}$ to increase with time according to \cref{eq:trunc_size}. This would have the complication of constantly adding new variables and defining their initial condition for a slight gain in speed. We avoid this to keep a cleaner code structure. 

We have also included a \textit{conservative truncation scheme}, where $n_{\textrm{max}}$ and $\ell_{\textrm{max}}$ are set by \cref{eqn:conservative_trunc} at \textit{all} scales, i.e.\ $\ell_{\textrm{max}} = 10$ and $n_{\textrm{max}}=19$ at the default \texttt{gbh\_FA\_trigger}$=15$. This scheme evaluates the neutrino transfer function more accurately at large scales at the expense of being marginally slower. Both schemes reach permille-level accuracy, so we set the aggressive scheme, which is faster with the \texttt{ndf15} evolver, as the default. In the code, this is controlled by the input parameter \colorbox{lightgray}{\texttt{gbh\_truncation\_scheme}}, which takes the values \colorbox{lightgray}{\texttt{aggressive}}  (the default) or \colorbox{lightgray}{\texttt{conservative}}.

The user can choose to increase \texttt{gbh\_FA\_trigger} in the input file to solve the GBH for longer and transition to the DFA later. This mainly improves the accuracy on small scales, where the matter power spectrum error stays below the permille level over a wider range of $k$ (see \cref{fig:P_error_vs_trigger}), but it increases the GBH system size and slows down the code (e.g.\ the runtime roughly doubles for \texttt{gbh\_FA\_trigger}$=30$, see \cref{fig:runtime_vs_trigger}). For large enough \texttt{gbh\_FA\_trigger}, the value of $n_{\textrm{max}}$ and $\ell_{\textrm{max}}$ eventually become so large that our table of $\omega_n$ will not be enough for the perturbation equations. In this case, the code resets \texttt{gbh\_use\_table}~$= 0$ and uses quadrature to compute all $\omega_n$. Furthermore, increasing $n_{\textrm{max}}$ and $\ell_{\textrm{max}}$ beyond a threshold makes the GBH numerically unstable, as already noted in \cite{BastosdeSennaNascimento:2021cej}. This threshold, and the effect of increasing \texttt{gbh\_FA\_trigger} on the accuracy and runtime of the code, are studied further in \cref{app:trigger}.

Note that \texttt{gbh\_FA\_trigger} does not directly translate into the corresponding \texttt{ncdm} parameter in \texttt{CLASS}, \texttt{ncdm\_fluid\_trigger\_tau\_over\_tau\_k} (default value 31), which switches to the fluid approximation once $k\tau > 31$ \cite{lesgourgues2011cosmic}. The GBH switch is instead set by $y = k\Tau$, and $\Tau \leq \tau$. While neutrinos are relativistic $\Tau \approx \tau$, so the GBH switches to the DFA at $k\tau \approx 15$, i.e.\ earlier than \texttt{ncdm}. After the nonrelativistic transition, $\Tau$ grows much more slowly than $\tau$, and the ordering is reversed for modes that reach the trigger late. For $M_{\nu} = 0.1\eV$, we find that \texttt{ncdm} switches first for $k \lesssim 0.01\,\textrm{Mpc}^{-1}$ (with the GBH never switching for $k \lesssim 0.009\,\textrm{Mpc}^{-1}$), while the GBH switches first at larger $k$, e.g.\ at $z \approx 2400$ rather than $z \approx 950$ for $k = 0.1\,\textrm{Mpc}^{-1}$. Small scales are therefore handed over to the fluid approximation earlier in the GBH than in \texttt{ncdm}.

Also note that although the DFA described in \cref{sec:fa} is selected by default via setting \colorbox{lightgray}{\texttt{gbh\_fluid\_approximation}=3}, one can still use the fluid approximation methods that are used by \texttt{ncdm} via selecting \colorbox{lightgray}{\texttt{gbh\_fluid\_approximation}=0, 1, 2} for Ma and Bertschinger \cite{Ma:1995ey}, Hu \cite{Hu:1998kj}, and \texttt{CLASS} \cite{lesgourgues2011cosmic} fluid approximations, respectively.

For setting the initial conditions, we use \cref{eq:gbh_ics}. Since neutrinos are initially relativistic, one may be tempted to use $\omega_n=1/3$ and simplify the initial conditions. This is equivalent to setting $q=\epsilon_q$ when integrating the initial conditions for $\Psi_\ell$, \cref{eq:super_horizon}, over the momentum space. However, this should be strictly avoided. In \texttt{CLASS}, the initial conformal time $\tau_i$ that is used to start evolving perturbation variables depends on $k$ and is larger for smaller $k$. At large scales, therefore, the initial conditions are set at a time when neutrinos are not fully relativistic anymore. Not using the exact $\omega_n$ at the initial conditions will therefore introduce an error that could grow to be at the percent level in the neutrino transfer functions.

\section{Testing the Generalized Boltzmann Hierarchy}
\label{sec:testing}
In this section, we compare our implementation of the GBH in \texttt{CLASS} to the standard \texttt{ncdm} implementation for massive neutrinos. We show that the GBH reproduces the same results as \texttt{ncdm} with comparable runtime and accuracy, by comparing the neutrino transfer functions, the matter, neutrino, and cross power spectra, and the runtime of the massive neutrino sector. 
We also demonstrate that the GBH provides unbiased posteriors on cosmological parameters, by running MCMC chains with the GBH and comparing the resulting posteriors to those obtained with \texttt{ncdm} in \cref{app:mcmc}. 
Throughout, we place the full neutrino mass in a single massive eigenstate, $m_{\nu}=M_{\nu}$, with the other two species massless, so that $N_{\rm eff}=3.044$.
All comparisons are made against a high precision reference: an \ncdm{} run with the
sampling tolerances of the automatic momentum quadrature and the integration tolerance tightened, more moments included, and with the fluid approximation
disabled so that the exact Boltzmann hierarchy is solved at all times and all
scales. This reference run uses the following input parameters:

\begin{lstlisting}
    ncdm_high_prec = dict(
        ncdm_fluid_approximation=3,
        ncdm_quadrature_strategy=0,
        l_max_ncdm=30,
        tol_ncdm_bg=1e-8,
        tol_ncdm_newtonian=1e-8,
        tol_ncdm_synchronous=1e-8,
        tol_perturbations_integration=1e-6
    )
\end{lstlisting}

\begin{figure}[t]
    \centering
    \includegraphics[width=\textwidth]{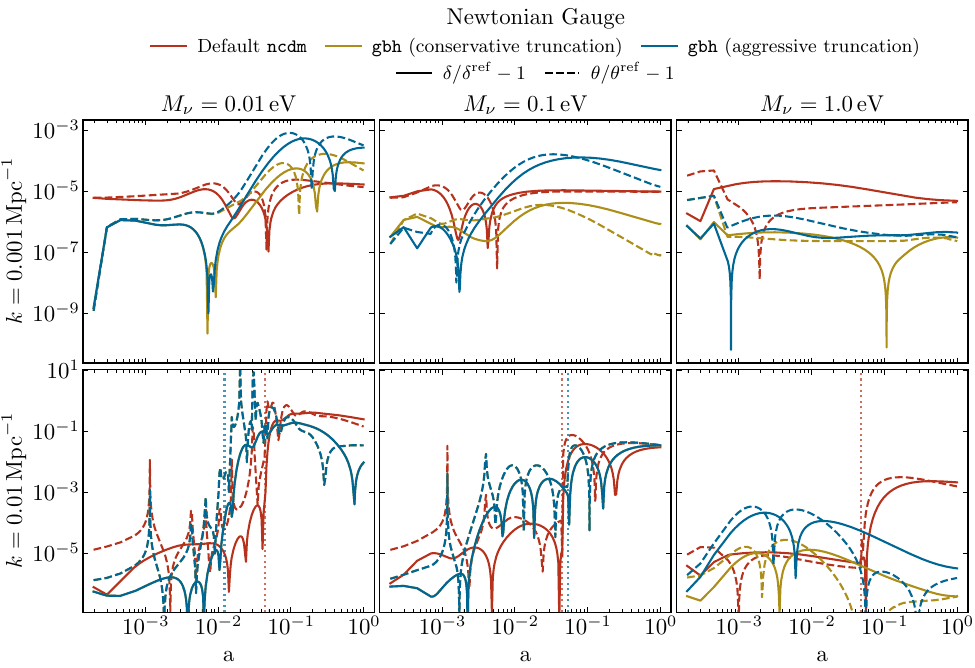}
    \caption{Fractional error in the neutrino density contrast $\delta$ (solid) and velocity divergence $\theta$ (dashed), in the Newtonian gauge, for both \ncdm{} (red) and \gbh{} (olive and blue), relative to a high-precision \ncdm{} reference run described in the text. 
    Columns compare different neutrino masses, while rows compare different comoving scales as labeled. 
    Results for GBH with both the aggressive [blue, \cref{eqn:nmax_lmax_vs_y}] and conservative [olive, \cref{eqn:conservative_trunc}] truncation schemes are shown.
    Dotted vertical lines mark the switch to a fluid approximation: blue for the \gbh{} switch to the DFA, which occurs at $y=k\Tau=15$ at the smaller scale ($k=0.01\,\textrm{Mpc}^{-1}$) for $M_\nu = 0.01\eV$, $0.1\eV$ and never for the largest mass, and red for the \ncdm{} switch to its fluid approximation at $k\tau = 31$, which occurs at the smaller scale for all three masses shown. 
    In the panels where the DFA is triggered, the two GBH truncation schemes coincide exactly and the olive curve is hidden behind the blue one.
    At the larger scale shown, $y(\tau_{0}) < 15$ at all three masses, so the GBH runs to $z=0$ with no switch, and the two truncation schemes differ: the aggressive scheme uses a smaller system and is correspondingly less accurate on these scales, by about an order of magnitude. 
    Errors are comparable to default \ncdm{} precision at all scales and masses, except for the smallest scale and largest mass, where the GBH is more accurate than the default fluid approximation used in \ncdm{}. While not shown, similar results hold in the synchronous gauge.
    }
    \label{fig:gbh_newt_perturb}
\end{figure}

\Cref{fig:gbh_newt_perturb} shows the fractional error in the neutrino
density and velocity divergence transfer functions, in the Newtonian gauge, for representative comoving scales and masses for both \ncdm{} and \gbh{} as compared to the reference cosmology. 
We compare  the ``aggressive'' and ``conservative'' GBH truncation schemes detailed in \cref{sec:implementation} separately.

At the largest scales, $y(\tau_{0}) = k\Tau(\tau_{0}) < $ \texttt{gbh\_FA\_trigger}, so the DFA is never triggered.
The full GBH is solved to present times, and the two truncation schemes differ. The conservative truncation scheme is more accurate, but the aggressive scheme, which is the default, still provides permille accuracy at all times. At smaller scales and lower masses, the two truncation schemes are identical when $y(\tau_{0}) > 15$ because they both use
\cref{eqn:conservative_trunc}.
Therefore, scales at which the two truncation schemes are identical coincide with modes that also necessarily transition to the DFA.
At larger masses but smaller scales, the GBH greatly outperforms the default \ncdm{} implementation, since the default fluid approximation in \class{} is inaccurate in this regime.
Regardless of the truncation scheme, scale, or mass, the GBH reproduces the reference transfer functions to permille accuracy at all times before the DFA is triggered. While not shown explicitly here, we have checked that similar results hold in the synchronous gauge.

After the DFA is triggered, the GBH reproduces the reference neutrino transfer functions to $10$--$30\%$ accuracy, which is comparable to the default \ncdm{} implementation at the scales shown. We emphasise that this is the expected and intended behaviour: these are modes where $\delta_{\nu}$ and $\theta_{\nu}$ are strongly suppressed and the DFA is deliberately approximate. These errors do not impact relevant observables, which we demonstrate below. The comparison with default \ncdm{} changes at larger $k$, beyond the scales in \cref{fig:gbh_newt_perturb}: the GBH error stays at the $\sim 10\%$ level, while that of the default \ncdm{} fluid approximation grows to order unity. The DFA tends to the correct asymptotic solution deep inside the free-streaming horizon. 
The neutrino auto power spectra show this improved small-scale modeling most clearly, as we now discuss.

\begin{figure}[t]
    \centering
    \includegraphics[width=\textwidth]{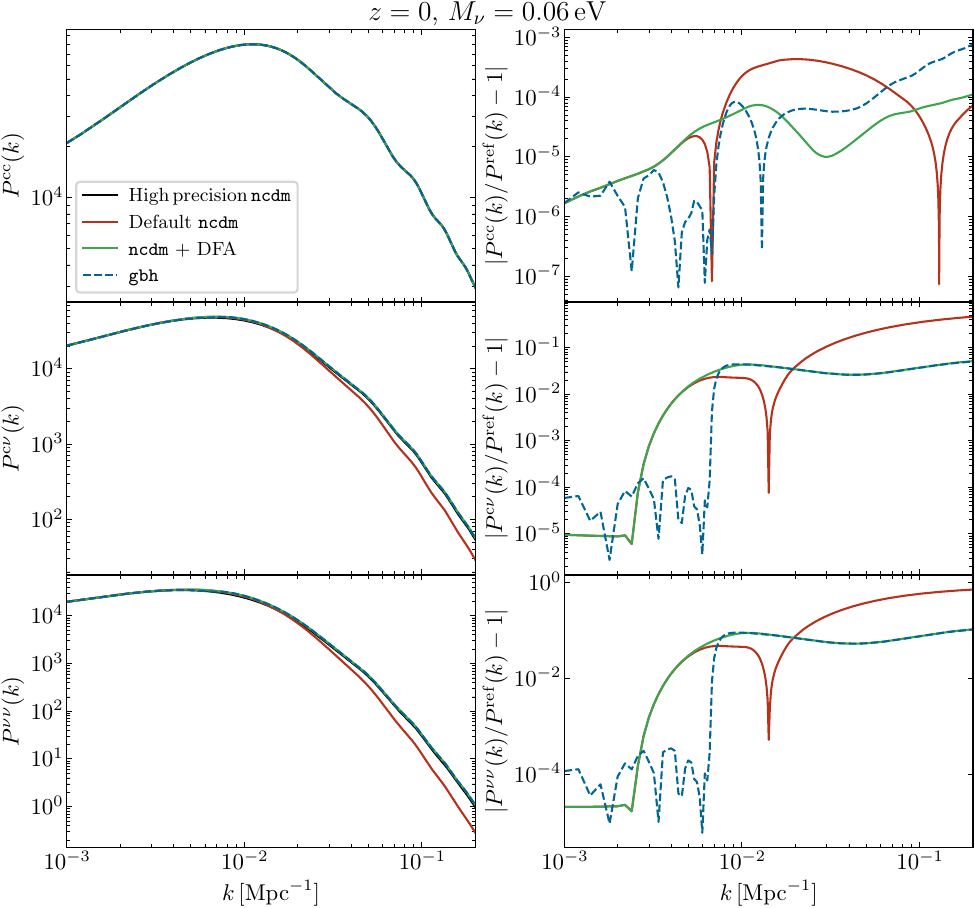}
    \caption{Comparisons of the cold dark matter autopower spectrum (top row, $P^{\rm cc}$), neutrino--cold dark matter cross spectra (middle row, $P^{\mathrm{c}\nu}$), and neutrino power spectrum (bottom row, $P^{\nu\nu}$) for \gbh{} and \ncdm{} species as labeled, at $z=0$ for $M_{\nu}=0.06\eV$. On the right, the fractional residuals for default \ncdm{} (red), \ncdm{} implemented with the DFA (green), and \gbh{} with default settings (blue, dashed) compared to a high precision \ncdm{} reference (black) are shown. While \gbh{} leads to larger errors in $P^{\rm cc}$ than both \ncdm{} configurations at small scales, $k \gtrsim 0.1\,{\rm Mpc}^{-1}$ (see text for details), all three agree with the reference to better than $10^{-3}$ out to $k=0.2$ \textrm{Mpc}$^{-1}$. 
    The \gbh{} and \ncdm{} with DFA spectra agree better with the reference than default \ncdm{} in the cross and neutrino power spectra, as the DFA significantly outperforms the default \class{} fluid approximation.}
    \label{fig:gbh_spectra}
\end{figure}

\Cref{fig:gbh_spectra} shows three spectra at $z=0$ for $M_{\nu} = 0.06\eV$: the cold dark matter auto power spectrum $P^{\rm cc}$, the neutrino auto power spectrum $P^{\nu\nu}$, and their cross spectrum $P^{{\rm c}\nu}$.  The left column shows the spectra, and the right column shows their fractional residuals relative to the high-precision \ncdm{} reference. We show the results for default \ncdm{} with the current fluid approximation in \class{}, the default \gbh{} configuration (aggressive truncation scheme and the DFA), as well as \ncdm{} but with the DFA.
The \ncdm{}$+$DFA and \gbh{} curves are
indistinguishable in $P^{{\rm c}\nu}$ and $P^{\nu\nu}$ for every mode that
reaches the fluid regime at this mass and trigger value. In fact, the DFA greatly improves on the default \class{} fluid approximation's precision and we advocate for its adoption as the default option for \ncdm{}. 

In the cold dark matter spectrum all three configurations agree with the reference to better than $10^{-3}$ across the full range, well within the
sensitivity of any current or planned survey. In particular, \ncdm{}$+$DFA keeps the error at or below the $10^{-4}$ level over the full range, improving on default \ncdm{} by up to an order of magnitude for $0.01 \lesssim k \lesssim 0.1\,{\rm Mpc}^{-1}$. However, \gbh{} is the most accurate at large scales but the least accurate at small scales, $k \gtrsim 0.1\,{\rm Mpc}^{-1}$, where it reaches $\approx 6\times10^{-4}$. 
This residual in $P^{\rm cc}$ that grows towards small scales is a consequence of the fluid approximation (which works best for nonrelativistic modes) taking over at early times for high-$k$ modes, while the neutrinos are still relativistic and still contribute appreciably to the gravitational potential. The resulting error is imprinted on the potential and propagates forward to
$z=0$. It can be reduced by raising \texttt{gbh\_FA\_trigger}, at a
computational cost, as we show in \cref{app:trigger}.
Since \ncdm{}+DFA uses the same fluid equations but switches later (see \cref{sec:implementation}), its difference when compared to the \gbh{} species is due to the timing of the switch rather than to the DFA itself.

\begin{figure}[t]
    \centering
    \includegraphics[width=\textwidth]{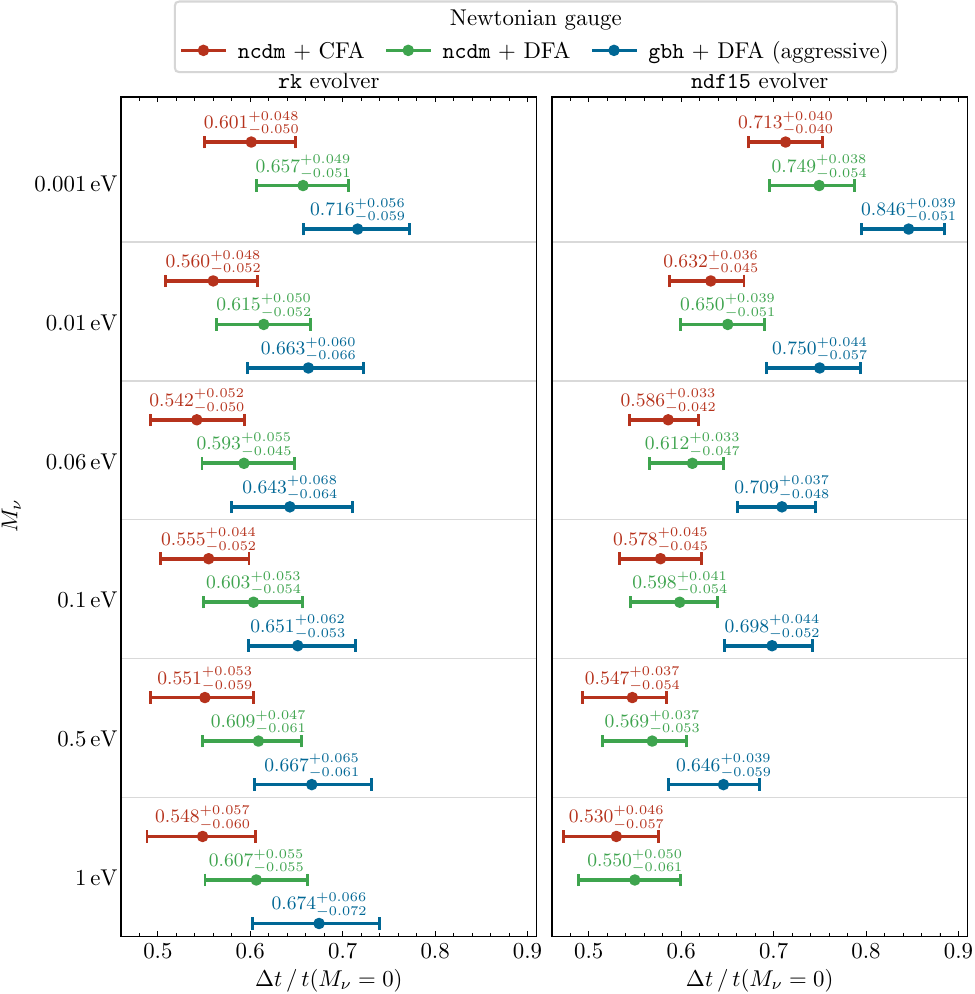}
    \caption{Fractional massive neutrino runtime overhead $\Delta t/t(M_{\nu}=0) = [t(M_{\nu})-t(M_{\nu}=0)]/t(M_{\nu}=0)$ in the Newtonian gauge, as a function of neutrino mass, for \ncdm{} with the current \class{} fluid approximation (CFA, red), \ncdm{} with the DFA (green), and the default \gbh{} configuration (aggressive truncation scheme and the DFA, blue), using the \texttt{rk} (left) and default \texttt{ndf15} (right) evolvers.
    Each point shows the median of 400 runtimes, marginalized over 20 randomly drawn cosmologies with $M_{\nu}$ fixed as labeled, with error bars and labels giving the 1$\sigma$ quantiles. Replacing the CFA by the DFA in \ncdm{} adds $0.02$--$0.06$ to the fractional overhead, i.e.\ only $1$--$4\%$ in total runtime. For both evolvers, \gbh{} spends $\approx 1.2$ times as long as \ncdm{} with the CFA on the massive neutrino sector, corresponding to $6$--$8\%$ in total runtime. No \gbh{} point is shown at $M_{\nu}=1\eV$ for \texttt{ndf15}, as the code switches to \texttt{rk} above $0.5\eV$ (see text for more details). Synchronous-gauge results, while not shown, exhibit similar behavior.
    }
    \label{fig:timing_plot}
\end{figure}

Finally, in \cref{fig:timing_plot}, we compare the computational cost of \gbh{}, default \ncdm{}, and \ncdm{} with the DFA across six neutrino masses and both ODE evolvers available in \class{}. 
We report the fractional
massive-neutrino overhead, i.e.\ the overhead normalised to the massless runtime,
$\Delta t / t(M_{\nu}=0) = [t(M_{\nu}) - t(M_{\nu}=0)]/t(M_{\nu}=0)$, which isolates the cost of the massive neutrino sector from the fixed cost of the rest of the Boltzmann solver. We use identical computational resources for all runs (40 cores on a single node at the University of Washington's \texttt{hyak} supercomputer) and report the distribution of 400 runtimes with $M_{\nu}$ fixed as labeled. More specifically, we draw 20 random cosmologies from a \Planck{}-compatible prior and run each 20 times: the repeated runs average over run-to-run timing noise, while the 20 cosmologies marginalize over the dependence of the neutrino overhead on the assumed cosmology.
We compare the effect of the two ODE evolvers available in \class{}: the default \texttt{ndf15} and the \texttt{rk} Runge--Kutta solver.

With the \texttt{rk} evolver, the fractional overhead is roughly independent of mass over three decades in $M_{\nu}$ for all three configurations: $0.54$--$0.60$ for \ncdm{} with the current \texttt{CLASS} fluid approximation (CFA), $0.59$--$0.66$ for \ncdm{} with the DFA, and $0.64$--$0.72$ for \gbh{}. With \texttt{ndf15}, the fractional overhead is larger and decreases with mass for all three, from $0.71$, $0.75$ and $0.85$ at $M_{\nu}=0.001\eV$ to $0.55$, $0.57$ and $0.65$ at $M_{\nu}=0.5\eV$, respectively (see \cref{sec:iterative} for a discussion of the evolver dependence).
The \texttt{ndf15} solver employs an adaptive step solver that incurs higher computational cost at light masses since modes remain relativistic for longer, requiring a smaller step size to resolve their oscillations.
On the other hand, with the \texttt{rk} solver the step is instead a fixed fraction of $\min(1/k,\,a/a')$ throughout the evolution, so all three scenarios take the same number of steps (to better than $1\%$), with a much weaker dependence on mass.

For both evolvers, switching \ncdm{} from the CFA to the DFA costs only $1$--$4\%$ in total runtime, and \gbh{} is a further $3$--$6\%$ slower than \ncdm{} with the DFA, i.e.\ $6$--$8\%$ slower in total than default \ncdm{}.
The DFA is slightly slower than the CFA due to evaluating the scale-dependent sound speed of \cref{eq:cs_fa_trunc}, but we support its adoption as the default option for \ncdm{} in \class{} due to the increase in accuracy for relatively small computational cost.
Despite the \gbh{} species requiring no momentum integrals in solving the perturbations while maintaining a similar system size as the \ncdm{} species, evaluating the right-hand side of \cref{eq:gbh_new} takes longer than the required computational cost for \ncdm{}.
Therefore the \gbh{} remains slightly slower across all masses, regardless of solver. 

The \texttt{ndf15} solver can fail to integrate the GBH at large masses, either crashing or causing large errors in the power spectrum.
For our fiducial cosmology this only happens above $M_{\nu}\approx 2\eV$, but for some parameter variations the onset is already at $M_{\nu}\approx 1\eV$.
Since the failure can be silent, the code conservatively switches to the \texttt{rk} solver automatically for $M_{\nu}>0.5\eV$.\footnote{Note that if one also turns off the tight-coupling-approximation, the \texttt{rk} solver will struggle to converge.}
For this reason, the right-hand panel of \cref{fig:timing_plot} shows \gbh{} timings only up to $M_{\nu}=0.5\eV$.
The failure is specific to the exact GBH hierarchy, and most likely arises because the adaptive step-size control of \texttt{ndf15} accepts steps too large to resolve the stiff dynamics of the hierarchy, so that the resulting error grows into a spurious instability.
We leave a fix to future work (see \cref{sec:iterative} for one possible approach).

\section{Future directions for the GBH}
\label{sec:generalization}

In \cref{sec:testing} we showed that the GBH, combined with the DFA, reproduces the standard \texttt{CLASS} \texttt{ncdm} predictions for massive neutrinos with the time spent on massive neutrinos $\sim 1.2$ times that of \texttt{ncdm}, corresponding to only a $\sim 7\%$ increase in total runtime (see \cref{fig:timing_plot}). All the while, the GBH relies on a substantially simpler perturbation-level implementation. These results were obtained for massive neutrinos specifically, both because they are the noncold relic species of greatest current observational interest and because their well-understood physics offers a controlled setting to validate the method. However, neither the GBH formalism of \cref{sec:theory} nor the numerical strategy of \cref{sec:implementation} makes any assumption specific to neutrinos, which raises the question of how broadly this approach can be expected to pay off.

The derivation of \cref{sec:theory} does rest on three ingredients that need not hold for other noncold relic species: (i) the background distribution function depends on comoving momentum $q$ alone; (ii) the Vlasov equation is collisionless, with no source or scattering term on the right-hand side of the perturbed Boltzmann equation; and (iii), as discussed in \cref{sec:applications}, the moment hierarchy is closed by a truncation scheme that may be model dependent, just as the truncation used by any fluid approximation of a noncold relic species may need to be. Each of these three ingredients can in principle be relaxed, but only on a case-by-case basis (see e.g. Ref.~\cite{Albertini:2024yfm}), requiring a dedicated derivation for each new collision or source term rather than following automatically from the formalism of \cref{sec:theory}. We leave a more general treatment to future work.

In the remainder of this section we consider two directions for extending the reach of the GBH beyond the present application. First, in \cref{sec:applications} we discuss under what conditions the GBH may offer a simpler and/or more efficient alternative to the full Boltzmann hierarchy for noncold relic species other than massive neutrinos, and argue that the most promising targets are species whose $f_{0}(q)$ is smooth but whose shape is not fixed once and for all. We illustrate this idea with a non-thermal toy model in \cref{sec:toy_model}. Second, in \cref{sec:iterative} we propose an iterative approach to solving the GBH system of equations that could further improve its efficiency, applicable to massive neutrinos as well as to any other noncold relic species.

\subsection{Applications to other noncold relics}
\label{sec:applications}

Beyond massive neutrinos, we expect the GBH to be most useful for noncold relics whose $f_{0}(q)$ is smooth but expensive to sample on a momentum grid, because its shape evolves in time or varies with the model parameters. Examples include freeze-in dark matter produced through a heavy mediator, non-resonantly produced sterile neutrinos in non-standard cosmologies, and decaying massive neutrinos. Below we explain where this cost comes from, discuss these and other concrete model examples, and explain why the GBH may be less suited to distributions that are not smooth.

The standard Boltzmann hierarchy for a noncold relic species requires solving separate multipole equations for each point on a discretized momentum grid, so its cost scales directly with the number of momentum points needed to resolve the momentum dependence of the perturbations, $\Psi_{\ell}(q)$. In \texttt{CLASS}, this cost is kept under control for massive neutrinos by an automatic quadrature strategy that combines an adaptive Gauss-Kronrod rule with a Gauss-Laguerre rule tailored to the smooth, thermal Fermi-Dirac shape of \cref{eq:Fermi-Dirac}. Because Laguerre quadrature converges quickly for such a distribution, as few as $5$ momentum points (at default precision settings) are needed at the perturbation level to reach the required accuracy. This efficiency is therefore tied to the assumed thermal shape of $f_{0}(q)$: for a noncold relic species with a distorted or non-thermal momentum distribution, the Laguerre rule is no longer guaranteed to be a good fit, the automatic strategy falls back on the more general adaptive quadrature, and correspondingly more momentum points may be needed to reach convergence.

This is not merely a hypothetical concern, and a number of well-motivated noncold relic candidates in the literature fall precisely into this regime. Decaying dark matter illustrates how severe this cost can become: unlike massive neutrinos, whose relic distribution is set once and for all at decoupling, the decay products of a dark matter particle are sourced continuously over cosmic time~\cite{Aoyama:2014tga,Kaplinghat:2005sy,Poulin:2016nat}, giving rise to a considerably more complex momentum distribution. In practice, solving the resulting Boltzmann hierarchy at the precision required for CMB and matter power spectrum analyses has been found to require $\mathcal{O}(10^{4})$ momentum bins, i.e. three orders of magnitude more than the handful sufficient for massive neutrinos at default precision. This amounts to $\mathcal{O}(10^{8})$ coupled differential equations and CPU times of a day or more per parameter choice, which is computationally prohibitive for a systematic scan of the model parameter space~\cite{FrancoAbellan:2021sxk,FrancoAbellan:2020xnr,Simon:2022ftd,Fuss:2022zyt,DEramo:2025jsb}.\footnote{The integral-equation approach of Ref.~\cite{Lee:2026zrl} reduces this to $\mathcal{O}(10^{2})$ momentum bins for the perturbations.}

Freeze-in dark matter provides another example. For a sufficiently light mediator, the resulting phase-space distribution can be so far out of thermal equilibrium that no fluid description applies. A full Boltzmann treatment is then required, implemented in \texttt{CLASS} by replacing the Fermi-Dirac distribution with a non-thermal distribution function~\cite{Dvorkin:2020xga,DEramo:2020gpr,Decant:2021mhj,Kamada:2019kpe}. More broadly, Ref.~\cite{Ballesteros:2020adh} surveys a wide range of related non-thermal production mechanisms, including the decay of inflaton or moduli condensates, the decay of thermalized or non-thermalized particles, and thermal or non-thermal freeze-in, with applications to gravitino, $Z'$- and spin-2-mediated, and spin-3/2 dark matter. The authors note that the resulting phase-space distributions can become sufficiently non-smooth that \texttt{CLASS}'s standard machinery, built around distributions like \cref{eq:Fermi-Dirac} that decay smoothly at large $q$, is no longer guaranteed to handle them robustly.

Beyond decaying and freeze-in dark matter, several other well-motivated candidates share the same basic feature: a distorted, non-thermal $f_{0}(q)$ handled through the generic momentum-grid machinery of Boltzmann codes for custom phase-space distributions, such as the \texttt{ncdm} module of \texttt{CLASS} or the massive-neutrino module of \texttt{CAMB}. These include non-thermal hot dark matter such as Dodelson-Widrow-like sterile neutrinos and hidden-sector species~\cite{Das:2021pof,Acero:2008rh,daCosta:2023mow,Bianchini:2023ubu}, and dark matter produced by the evaporation of primordial black holes~\cite{Chen:2025gnm,Baldes:2020nuv,Shallue:2024hqe}. Closer to the regime outlined at the start of this subsection are species whose $f_{0}(q)$ remains smooth, i.e.\ unimodal with no sharp features or discontinuities, but whose shape is not fixed once and for all as it is for massive neutrinos.

Three examples from the literature illustrate this case well. For freeze-in dark matter produced through a heavy s-channel mediator, Ref.~\cite{Frangipane:2021rtf} finds a momentum distribution whose peak sits at $p/T \sim m_{h}/(2T_{\mathrm{RH}})$, and both the peak location and the width of the distribution shift continuously as the reheating temperature $T_{\mathrm{RH}}$ or the mediator mass $m_{h}$ are varied. Similarly, for non-resonantly produced sterile neutrino dark matter, Ref.~\cite{Gelmini:2019wfp} derives a closed-form $f_{0}(q)$ in non-standard pre-BBN cosmologies where the expansion rate is modified as $H = \eta\,(T/T_{\rm tr})^{\beta} H_{\rm Std}$. The result is a Fermi-Dirac distribution multiplied by a power law, $f_{0} \propto (q/T)^{\beta/3}$, so the shape is set by the index $\beta$ of the expansion history: it reduces to the Fermi-Dirac form in the standard cosmology and becomes warmer ($\beta>0$) or colder ($\beta<0$) otherwise, while the sterile neutrino mass and mixing angle only set its normalization. A third example is decaying or unstable massive neutrinos~\cite{Chacko:2019nej,FrancoAbellan:2021hdb,Barenboim:2020vrr,Holm:2022eqq}, where the momentum distribution of the parent population retains its Fermi-Dirac form multiplied by a decay factor $\exp(-\Gamma t/\gamma)$. Since $\gamma = E/m$ depends on $q$, this suppression is momentum-dependent, so the shape itself continuously drifts away from thermal as time goes on, at a rate set by the free parameter $\Gamma$ (the neutrino lifetime).

In the first two examples, the shape of $f_{0}$ changes across parameter space but not in time. This poses no problem of principle for \ncdm{}, whose automatic quadrature strategy selects a new momentum grid for each set of parameters by requiring that a test integral of $f_{0}$ converge. The cost of each evaluation, however, grows with the number of momentum points this requires, which in some models may become large in parts of the parameter space. In the toy model of \cref{sec:toy_model}, for instance, it grows from $6$ for a thermal distribution to $245$ for the broadest one we consider. The third example is more demanding. There the shape of $f_{0}$ drifts in time within a single run, but the momentum grid is chosen once, before the perturbations are evolved, so it must contain conservatively many momentum points to resolve $f_{0}$ at all times. 

We expect the GBH to be more efficient in both cases, and the toy model confirms this for a shape that changes across parameter space.\footnote{As is, the GBH set-up of \cref{sec:gbh} assumes a time-independent $f_{0}$, so it needs to be extended to allow for a time dependence, which we expect to be a technical rather than a conceptual complication (and has been done in one concrete application of the method \cite{Albertini:2024yfm}).} The reason is that observables only require momentum integrals of the perturbations, such as the density, pressure, velocity and shear perturbations. On the other hand, \ncdm{} must sample $\Psi_{\ell}(q)$ finely enough in $q$ to compute these integrals accurately, so its grid grows as the distribution becomes harder to sample. The GBH instead evolves the integrals directly, and its cost is set by how many moments the perturbations require. This need not track the shape of $f_{0}$, as we find in \cref{sec:toy_model}, although a distribution with a substantial high-momentum tail may still require a larger $n_{\textrm{max}}$, since faster particles free-stream further.\footnote{In our current implementation, the fitting function of \cref{eq:trunc_size} is calibrated for massive neutrinos and may need to be recalibrated for other species. The scale that sets $n_{\textrm{max}}$ for each mode, on the other hand, can already be adapted to a given distribution function through \texttt{gbh\_q\_horizon} (see \cref{sec:implementation}), and is enough to ensure convergence in the toy model of \cref{sec:toy_model}.}

However, the GBH does not obviously help with distributions that are not smooth, such as some of those surveyed in Ref.~\cite{Ballesteros:2020adh}, and may in fact make matters worse. \Cref{eq:gbh_variables} defines the $n$-moments as weighted integrals directly over $f_{0}(q)$. So a sharply-featured, multi-peaked, or discontinuous $f_{0}(q)$ (e.g. a resonantly-produced~\cite{Baur:2017stq,Cherry:2017dwu,Boyarsky:2008mt,Ghiglieri:2015jua,Venumadhav:2015pla} or narrowly-sourced species~\cite{Garcia:2022vwm,Kaplinghat:2005sy}) generically requires a large $n_{\textrm{max}}$ to resolve, for the same reason any moment expansion struggles with non-smooth functions. 

On top of this, the truncation closure of \cref{eq:trunc_nmax} is not just a statement about how many moments are needed: it is derived (see footnote~\ref{fn:trunc_nmax}) assuming $\Psi_{0} \propto d\log f_{0}/d\log q$ and $\Psi_{\ell} \propto (q/\epsilon_{q})^{\ell-2} d\log f_{0}/d\log q$. For a sharp or non-smooth $f_{0}(q)$, $d\log f_{0}/d\log q$ can itself become large or ill-behaved near the feature, threatening the validity of the closure relation itself and not merely the required truncation order. In short, the GBH is not obviously advantageous over the standard hierarchy for such species, and may face a harder version of the same problem, namely closure validity rather than just cost.

A further caveat is that the GBH must eventually hand over to a fluid approximation at small scales, since the hierarchy becomes increasingly expensive and eventually numerically unstable as $k\Tau$ grows. As shown in \cref{app:trigger}, this switch can be delayed well beyond our default choice, at the price of a steadily growing runtime. In that case, the fluid approximation only needs to be trusted on correspondingly smaller scales. The GBH fails to help only when a species' fluid description remains inaccurate even on those scales, and for which accurate predictions are needed there. In such a case, the Boltzmann equation must be solved throughout.

A possible example is freeze-in dark matter with a sufficiently light mediator. Ref.~\cite{Dvorkin:2020xga} finds that the free-streaming suppression of the non-thermal phase-space distribution cannot be captured by fluid equations at all, and instead solves the full Boltzmann hierarchy. Similarly, Ref.~\cite{Decant:2021mhj} finds that the default \texttt{ncdm} fluid approximation in \texttt{CLASS} must be switched off to accurately compute the matter power spectrum of freeze-in dark matter on Lyman-$\alpha$ scales, which substantially slows down the computation. Another example is decaying dark matter. Ref.~\cite{FrancoAbellan:2021sxk} describes the warm decay products with a fluid approximation based on a fitting formula. This formula cannot capture the oscillatory features sourced by the free-streaming of decay products produced at different times. Even so, they find it accurate enough for the observables considered in that work. Ref.~\cite{Lee:2026zrl} subsequently quantifies this residual directly, using an independent integral-equation approach (the \classier{} method). They find percent-level deviations that grow on small scales. Whether a delayed switch reaches far enough for such species would need to be assessed on a case by case basis.

Taken together, these considerations suggest a class of noncold relic species for which the GBH has the potential to offer an advantage beyond massive neutrinos: species with a smooth background distribution whose shape is not fixed once and for all but instead drifts continuously across the model's parameter space or in time (e.g. the freeze-in, sterile neutrino, and decaying-neutrino examples of Refs.~\cite{Frangipane:2021rtf,Gelmini:2019wfp,Chacko:2019nej,FrancoAbellan:2021hdb,Barenboim:2020vrr,Holm:2022eqq} discussed above, and the time-dependent case already treated with the GBH in \citep{Albertini:2024yfm}). In addition, a small-scale fluid approximation must eventually become trustworthy, so that the GBH$+$fluid-approximation strategy of this paper remains viable. Outside this class, the standard \texttt{ncdm} machinery or the iterative, integral-equation-based \classier{} method \cite{Kamionkowski:2021njk,Ji:2022iji,Lee:2025zym,Lee:2026zrl} may be the more suitable choice.

\subsection{A non-thermal toy model}
\label{sec:toy_model}

\begin{figure}[t]
    \centering
    \includegraphics[width=0.6\textwidth]{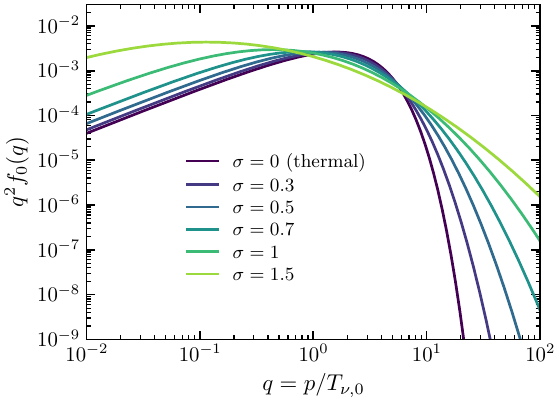}
    \caption{Distribution functions of the smeared Bose-Einstein toy model of \cref{eq:smeared_be}, plotted as $q^{2}f_{0}(q)$, for the values of $\sigma$ used in \cref{fig:err_timing_smeared}. Larger $\sigma$ broadens the distribution towards both low and high momenta at fixed number density, so that the \ncdm{} optimal quadrature strategy needs a larger number of momentum points to sample it accurately.}
    \label{fig:psd_smeared}
\end{figure}

As argued in \cref{sec:applications}, the GBH should be advantageous over \ncdm{} whenever accurately sampling a smooth $f_{0}$ on a momentum grid is expensive, since the GBH evolves momentum integrated quantities directly and needs no such grid for the perturbations. Ref. \citep{Albertini:2024yfm} has already applied the GBH to a distribution that broadens in time. We now illustrate this with a toy model in which a single parameter, $\sigma$, controls the broadening of a thermal distribution. We start from a Bose-Einstein distribution and smear its temperature $T$ over a lognormal distribution of width $\sigma$ around $T_{0}$,
\begin{equation}
\label{eq:smeared_be}
    f_{0}(q) = \frac{g_{s}}{(2\pi)^{3}} \int_{-\infty}^{\infty} d\ln T \, \frac{e^{-\ln^{2}(T/T_{0})/2\sigma^{2}}}{\sqrt{2\pi}\,\sigma} \left(\frac{T_{0}}{T}\right)^{3} \frac{1}{e^{q/T}-1} \,,
\end{equation}
with $g_{s}=1$ and $T_{0}$ a reference temperature, which we set to the present-day neutrino temperature. The factor $(T_{0}/T)^{3}$ keeps the number density independent of $\sigma$, and $\sigma=0$ recovers the standard thermal Bose-Einstein distribution. As shown in \cref{fig:psd_smeared}, increasing $\sigma$ broadens $f_{0}(q)$ towards both low and high momenta while keeping it smooth and unimodal. This is a smooth, non-thermal distortion of a thermal spectrum, of the kind discussed in \cref{sec:applications}, without the details of any particular production mechanism.\footnote{We use Bose-Einstein rather than Fermi-Dirac statistics because smearing a Fermi-Dirac distribution in this way raises its occupation number per state above unity at low momentum, which the Pauli exclusion principle forbids.}

We fix the abundance to that of a single massive neutrino species with $M_{\nu}=0.06\eV$, i.e.\ a physical density $\Omega_{\textrm{ncdm}} h^{2} = 6.44\times10^{-4}$, which sets the particle mass to $m \approx 0.090\eV$ for all $\sigma$. As in \cref{sec:testing}, the other two neutrino species are massless, all other cosmological parameters are fixed, and we work in the Newtonian gauge. We compute the matter power spectrum at $z=0$ with default \ncdm{} (with the \class{} fluid approximation) and with default \gbh{} (aggressive truncation scheme and the DFA), except for \texttt{gbh\_q\_horizon = auto} (see \cref{sec:implementation}). Both codes read the same table of $f_{0}(q)$, obtained by evaluating \cref{eq:smeared_be} on a fine logarithmic grid in $q/T_{0}$.

\begin{figure}[t]
    \centering
    \includegraphics[width=0.6\textwidth]{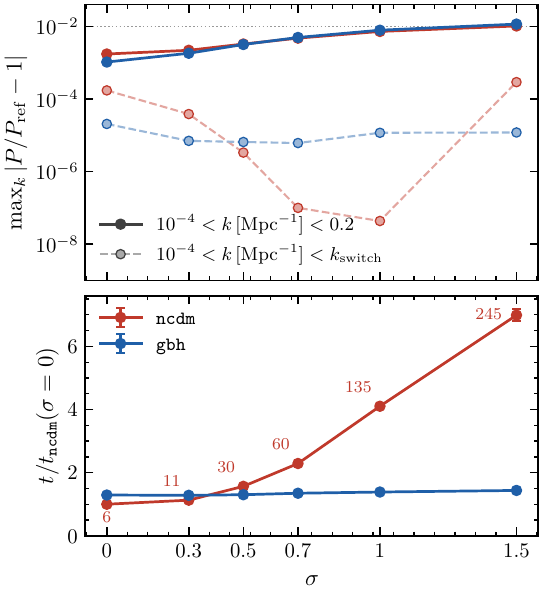}
    \caption{Accuracy (top) and runtime (bottom) of \ncdm{} (red) and \gbh{} (blue) for the smeared Bose-Einstein toy model of \cref{eq:smeared_be}, as functions of the broadening parameter $\sigma$. Top: largest fractional error in the matter power spectrum at $z=0$, relative to the high-precision \ncdm{} reference, over $10^{-4}<k<0.2\,\textrm{Mpc}^{-1}$ (dark markers) and over $10^{-4}\,\textrm{Mpc}^{-1}<k<k_{\rm switch}$ (light markers). Here $k_{\rm switch}$ is the largest $k$ for which the fluid approximation never triggers, $k_{\rm switch}\approx 2.2\times10^{-3}\,\textrm{Mpc}^{-1}$ for \ncdm{} and between $8.5\times10^{-3}$ and $4.8\times10^{-3}\,\textrm{Mpc}^{-1}$ for \gbh{}, decreasing with $\sigma$. The dotted line marks a percent relative error. The light markers thus show the accuracy of the full Boltzmann hierarchies. Bottom: total runtime divided by that of \ncdm{} at $\sigma=0$, with mean and standard deviation over ten repetitions. Numbers next to the \ncdm{} points give its number of momentum points, selected by the automatic quadrature strategy of \class{}.}
    \label{fig:err_timing_smeared}
\end{figure}

\Cref{fig:err_timing_smeared} summarizes the accuracy and cost of both codes as functions of $\sigma$. In the top panel, the largest fractional error in the matter power spectrum over $10^{-4}<k<0.2\,\textrm{Mpc}^{-1}$ grows for both codes from $\approx 0.1\%$ at $\sigma=0$ to $\approx 1\%$ at $\sigma=1.5$. As in \cref{sec:testing}, this error comes from the fluid approximation. Disabling it in \ncdm{} while keeping the same momentum grid reduces the error to $ <0.1\%$ for all $\sigma$. Restricting to modes that never switch to a fluid approximation, $k<k_{\rm switch}$ (see \cref{app:trigger}), isolates the accuracy of the Boltzmann hierarchies themselves. There, the \gbh{} error stays of order $\sim 10^{-5}$ for all $\sigma$, with the same truncation rule throughout and no tuning to the shape of $f_{0}$. The \ncdm{} error over the same range varies between $10^{-7}$ and $3\times10^{-4}$, depending on how well its automatic momentum sampling happens to resolve each distribution.

The bottom panel shows the total runtime of each code, normalized to that of \ncdm{} at $\sigma=0$. The automatic quadrature of \ncdm{} picks $6$ momentum points at $\sigma=0$, comparable to the $5$ used for massive neutrinos, but $245$ at $\sigma=1.5$, and its runtime grows by a factor of $\sim 7$ accordingly. The runtime of \gbh{} instead grows only from $\approx 1.3$ to $\approx 1.4$ times that of thermal \ncdm{}, so that \gbh{} is already the faster of the two at $\sigma=0.5$, and $\sim 5$ times faster at $\sigma=1.5$. This is the behavior anticipated in \cref{sec:applications}. The cost of \ncdm{} follows the momentum grid needed to sample $f_{0}$, while that of \gbh{} is mostly set by the truncation orders of \cref{eq:trunc_size}, which depend on $f_{0}$ only through the horizon evaluated at $\langle q\rangle$.

The toy model is deliberately simple, but it nonetheless shows that, for a smooth distribution function whose shape varies across parameter space, a fixed truncation rule with the GBH can keep both accuracy and cost stable, while the cost of a momentum grid in the standard implementation grows as the distribution becomes harder to sample.

\subsection{An iterative approach to solving the GBH}
\label{sec:iterative}

\Cref{sec:testing} showed that the extra time \gbh{} spends on massive neutrinos, relative to a run without them, is larger under \texttt{CLASS}'s default evolver, \texttt{ndf15}, than under the explicit \texttt{rk} evolver. It is also somewhat larger than that of the standard \texttt{ncdm} module (see \cref{fig:timing_plot}). Moreover, \texttt{ndf15} becomes numerically unreliable for the GBH's exact hierarchy above $M_{\nu}\sim1$--$2\,\textrm{eV}$ (depending on cosmology). Both point to the same fix: rather than embedding the GBH in the generic, one-size-fits-all evolver used for the rest of the perturbation system, integrate it with a dedicated numerical strategy. Such a strategy can exploit the fact that the GBH equations form a large linear block whose coefficient matrix is known in closed form and has a simple sparse structure. The generic evolver, by contrast, treats them as part of a system whose Jacobian is estimated numerically by finite differences. We outline such a strategy below, inspired by the \classier{} method~\cite{Kamionkowski:2021njk,Ji:2022iji,Lee:2025zym}.

The starting point is to make the structure just described explicit. Stack the moments $f_{\ell,n}$ (with the conventions $f_{0,n}\equiv\delta_{n}$ and $\theta_{n}=kf_{1,n}$ introduced below \cref{eq:gbh_variables}) for $0\leq\ell\leq\ell_{\mathrm{max}}$ and $0\leq n\leq n_{\mathrm{max}}$ into a single vector $X(\tau)$ of dimension $N=(\ell_{\mathrm{max}}+1)(n_{\mathrm{max}}+1)$. In this notation, the truncated GBH system of \cref{eq:gbh_new,eq:gbh_sync} is simply
\begin{equation}
\label{eq:gbh_matrix}
    X'(\tau) = M(\tau) X(\tau) + S(\tau) \,,
\end{equation}
where $M(\tau)$ is the same coefficient matrix in both gauges, and only the source $S(\tau)$ depends on the gauge choice: it collects the $\{\phi',\psi\}$ metric-sourcing terms of \cref{eq:gbh_new} in the Newtonian gauge, or the $\{h',\eta'\}$ terms of \cref{eq:gbh_sync} in the synchronous gauge. These are nonzero only on the $\delta_{n}$ and $\theta_{n}$ entries of $X$ (and, in the synchronous gauge, the $\sigma_{n}$ entries as well, through the $\delta^{K}_{\ell,2}$ term). \Cref{eq:gbh_matrix} is not new content, but simply the GBH system of \cref{eq:gbh_new,eq:gbh_sync} in matrix notation. This form nonetheless isolates the structural facts that motivate what follows. First, $M(\tau)$ depends only on the background quantities $\mathcal{H}(\tau)$ and the equations of state $\omega_{n}(\tau)$, together with $k$, and never on $X$ itself. Second, $M(\tau)$ is sparse: reading off \cref{eq:gbh_new}, the equation of motion for each entry $f_{\ell,n}$ couples it to at most four others: itself, $f_{\ell,n+1}$, $f_{\ell-1,n+1}$, and $f_{\ell+1,n}$. So $M(\tau)$ is banded on the $(\ell,n)$ grid, with $\mathcal{O}(N)$ nonzero entries rather than the $N^{2}$ of a dense $N\times N$ matrix. In short, \cref{eq:gbh_matrix} shows that the truncated GBH is a linear, non-autonomous, sparse ODE system with an exactly known coefficient matrix --- a structure that \texttt{CLASS}'s evolvers, built to handle fully generic nonlinear systems, do not exploit.

The choice of numerical method for \cref{eq:gbh_matrix} matters, and neither of the two evolvers used for the GBH in \cref{sec:testing} is a good match to its structure. The explicit \texttt{rk} evolver is stability-limited by the eigenvalues of $M(\tau)$: the diagonal decay term $-(2n-3\omega)\mathcal{H}$ in the equation of motion for $\delta_{n}$ [see \cref{eq:gbh_new}] grows linearly with $n$, so the stiffest eigenvalues of $M(\tau)$ scale with $n_{\mathrm{max}}$. This forces an explicit method to take a step size small enough to remain stable against them, and this only gets worse as $n_{\mathrm{max}}$ grows with $k$. The implicit \texttt{ndf15} evolver avoids this stability restriction, but only by relying on a Jacobian estimated by finite differences, even though $M(\tau)$ is already known exactly. This is also the same evolver that, as noted above, becomes numerically unreliable for the GBH above $M_{\nu}\sim1$--$2\,\textrm{eV}$. Its adaptive step-size control either under-resolves the stiff GBH dynamics, giving incorrect output, or, at still larger masses, shrinks the step size until the integration aborts. For the full Einstein--Boltzmann system, Ref.~\cite{Sletmoen:2025fro} similarly finds backward differentiation formula methods such as \texttt{ndf15} to be the slowest of the implicit solvers tested. The most efficient was a Rosenbrock method using the exact, analytically generated sparse Jacobian (\texttt{Rodas5P}).

An exponential integrator avoids both problems at once. It treats the stiff linear part $M(\tau)X(\tau)$ of \cref{eq:gbh_matrix} through the matrix exponential $\exp[M(\tau)\Delta\tau]$~\cite{HochbruckOstermann:2010}, with $M(\tau)$ evaluated within each step. Since $M(\tau)$ is sparse, the action of this exponential on $X(\tau)$ can be computed cheaply with Krylov-subspace methods~\cite{Saad:1992,HochbruckLubich:1997,Sidje:1998}. Only the comparatively smooth source term needs to be approximated within each step. Because the matrix exponential propagates even the stiffest modes of $M(\tau)$, the scheme is unconditionally stable with respect to them, regardless of step size (unlike the \texttt{rk} evolver). It also uses the coefficient matrix directly rather than rediscovering it by finite differences (unlike the \texttt{ndf15} evolver). This remains a local, per-step operation on the current $X(\tau)$ alone, so it keeps the same $\mathcal{O}(N)$ per-step cost as the existing evolvers.

This fast per-step solver cannot simply replace \texttt{CLASS}'s existing evolver for the GBH alone. In \texttt{CLASS}, the GBH's degrees of freedom are packed together with those of every other species and the metric into a single joint state vector. This vector is evolved by one call to a generic stiff solver, so no species-specific solver can be substituted unless the GBH block is first decoupled from the rest of the system. The natural way to do so is an outer iteration: fix a metric history, solve the GBH block alone with the exponential integrator described above, use the resulting neutrino stress-energy to re-solve the rest of the system and obtain an updated metric history, and repeat until the metric history converges.

This is exactly the outer-loop structure that \classier{} already implements in \texttt{CLASS}'s perturbation module, albeit using a different technique for \texttt{ncdm}. It uses an integral equation for the phase-space distribution evaluated through convolutions in Fourier space, rather than the GBH's own linear ODE system. \classier{} obtains its initial metric history (the ``0th iteration'') from a default \texttt{CLASS} run, i.e. \texttt{ncdm}'s own low-$\ell_{\textrm{max}}$ Boltzmann hierarchy together with the standard fluid approximation. It then alternates between the integral-equation solver and the rest of the system as described above. Refs.~\cite{Lee:2025zym,Lee:2026zrl} find that this converges rapidly for both massive neutrinos and decaying dark matter, with at most two iterations typically sufficient for the required precision. We propose to reuse this same outer loop, taking the DFA of \cref{sec:fa} as the 0th-iteration ansatz for the metric history, but applying the exponential integrator to the GBH system of \cref{eq:gbh_matrix} at each subsequent iteration.

With the exponential integrator, each evaluation of the GBH block should avoid both the numerical unreliability of \texttt{ndf15} and the step-size limitation of \texttt{rk} identified above. The outer loop itself should add comparatively little cost: in \classier{}, re-integrating the rest of the system with the noncold-relic sector held fixed accounts for only $\sim 10$--$25\%$ of the total runtime (see Fig.~4 in \cite{Lee:2025vgv}). The dominant cost would then lie in solving the GBH within each iteration. Free-streaming makes the GBH solution oscillate with phase $\sim k\Tau$, i.e. on a timescale $\Tau \sim 1/k$, and the stiffness of $M(\tau)$ grows with $n_{\mathrm{max}}$. So the Krylov subspace needed per step, and hence the cost of the exponential integrator, is expected to grow with $k$. In practice this growth is capped by the switch to the DFA below the free-streaming scale (see \cref{app:trigger}). We leave a full implementation of this approach to future work.

\section{Conclusion}
\label{sec:conc}

Accurate predictions for noncold relic species in linear Boltzmann solvers require resolving the momentum dependence of their perturbations. In the standard approach, this means evolving the Boltzmann hierarchy on a momentum grid tailored to the background distribution function. In this work we presented the first open-source implementation in \texttt{CLASS} of an alternative approach, in which the momentum dependence is integrated out through the Generalized Boltzmann Hierarchy (GBH) of Ref.~\cite{BastosdeSennaNascimento:2021cej}. This approach trades the momentum grid for a hierarchy of higher velocity-weighted moments of the distribution function. In addition, small scales are handed over to the Dispersive Fluid Approximation (DFA) of Ref.~\cite{Nascimento:2023psl}. We also implemented the DFA in the standard \ncdm{} module, where it can replace the default \class{} fluid approximation independently of the GBH. We validated this implementation against the standard \ncdm{} module, using massive neutrinos as a benchmark. Our main findings are summarized below.

\begin{itemize}
    \item In \cref{sec:testing} we showed that the GBH reproduces the reference neutrino transfer functions to permille accuracy before the DFA is triggered, as shown in \cref{fig:gbh_newt_perturb}. \Cref{fig:gbh_spectra} shows that it also reproduces the cold dark matter power spectrum to better than $10^{-3}$ at all scales. The DFA significantly outperforms the default \class{} fluid approximation in the cross and neutrino power spectra, with errors in the latter $\lesssim 10\%$ at all scales. The GBH$+$DFA also yields cosmological parameter constraints in excellent agreement with \ncdm{}, with marginalized median values of all cosmological parameters agreeing to better than $0.05\sigma$ (see \cref{fig:mcmc} in \cref{app:mcmc}).
    \item Also in \cref{sec:testing}, we showed that \gbh{} comes at a computational cost comparable to that of default \ncdm{}: it spends $\approx 1.2$ times as long on the massive neutrino sector, i.e.\ $6$--$8\%$ more in total runtime, while adopting the DFA in \ncdm{} costs only $1$--$4\%$ (\cref{fig:timing_plot}). We therefore advocate for adopting the DFA in the standard \ncdm{} module, regardless of whether the GBH is also used. The GBH can become numerically unstable with \texttt{ndf15} at large masses, so the code automatically switches to \texttt{rk} above $M_{\nu}=0.5\eV$.
    \item In \cref{sec:generalization} we discussed future directions. In \cref{sec:applications}, we identified a class of noncold relics for which the GBH may offer an advantage over the standard hierarchy: species with a smooth background distribution function that is expensive to sample on a momentum grid (for instance because its shape drifts across parameter space or in time), and for which a small-scale fluid approximation eventually becomes trustworthy. In \cref{sec:toy_model}, we tested this on a toy model in which a single parameter broadens a thermal distribution. Both codes reproduce the matter power spectrum to about $1\%$ or better, but the runtime of \gbh{} stays within $1.3$--$1.4$ times that of thermal \ncdm{}, while that of \ncdm{} grows by up to a factor of $\sim 7$ as the broadening increases, reflecting the growing cost of sampling the distribution function on a momentum grid. In \cref{sec:iterative}, we outlined an iterative approach, combining an exponential integrator with an outer loop over the metric history, that should fix the \texttt{ndf15} instabilities found in \cref{sec:testing} and could make the GBH more efficient.
\end{itemize}

Noncold relics, from massive neutrinos to hot subcomponents of the dark matter, are a key target of upcoming cosmological surveys. These surveys will be sensitive to the scale dependence of the free-streaming suppression imprinted on structure growth. Turning these measurements into constraints requires predictions that are both accurate and efficient enough to scan the parameter space of each model. Making such scans feasible for species beyond the standard massive neutrinos is the main motivation for the GBH. Natural next steps are applications of the GBH to concrete non-thermal species, building on the toy model of \cref{sec:toy_model}, and an implementation of the iterative approach outlined in \cref{sec:iterative}. Accurate and efficient tools of this kind will be essential to extract the full information that upcoming surveys carry about the nature of dark matter.

\acknowledgements

MMS thanks Masha Baryakhtar, Marilena Loverde, and Zachary J. Weiner for many helpful discussions and invaluable suggestions throughout the completion of this work.
CBSN thanks Nils Schöneberg, Maria Archidiacono, Eiichiro Komatsu, Nanoom Lee, Marc Kamionkowski,  Cara Giovanetti, Thomas Bakx and Marco Bonici for valuable discussions. A. N. acknowledges the support of the Natural Sciences and Engineering Research Council of Canada (NSERC), funding reference numbers RGPIN-2023-04121, RGPIN-2024-03851, and RGPIN-2020-05569. AN also acknowledges support during the first year of this work from STFC grant ST/W006537/1 and from the President’s PhD Scholarship from Imperial College London. MMS and CBSN acknowledge support from the Department of Energy grant DE-SC0011637. Research at Perimeter Institute is supported in part by the Government of Canada through the Department of Innovation, Science and Economic Development and by the Province of Ontario through the Ministry of Colleges and Universities. This work was enabled, in part, by the use of advanced computational, storage, and networking infrastructure provided by the Hyak supercomputer system at the University of Washington~\cite{uwhyak}, which was supported by the UW Student Technology Fee, the UW Department of Physics, and the College of Arts and Sciences. We acknowledge the use of various large language models by Anthropic and OpenAI for refining plotting code, assisting in literature search, testing and improving code, and editing text. This work made use of a number of open-source software packages~\cite{Harris:2020xlr, Virtanen:2019joe, Hunter:2007ouj,hoyer2017xarray}.

\appendix

\section{The sub-horizon breakdown of the GBH}
\label[appendix]{app:trigger}

\begin{figure}[t]
    \centering
    \includegraphics[width=0.75\textwidth]{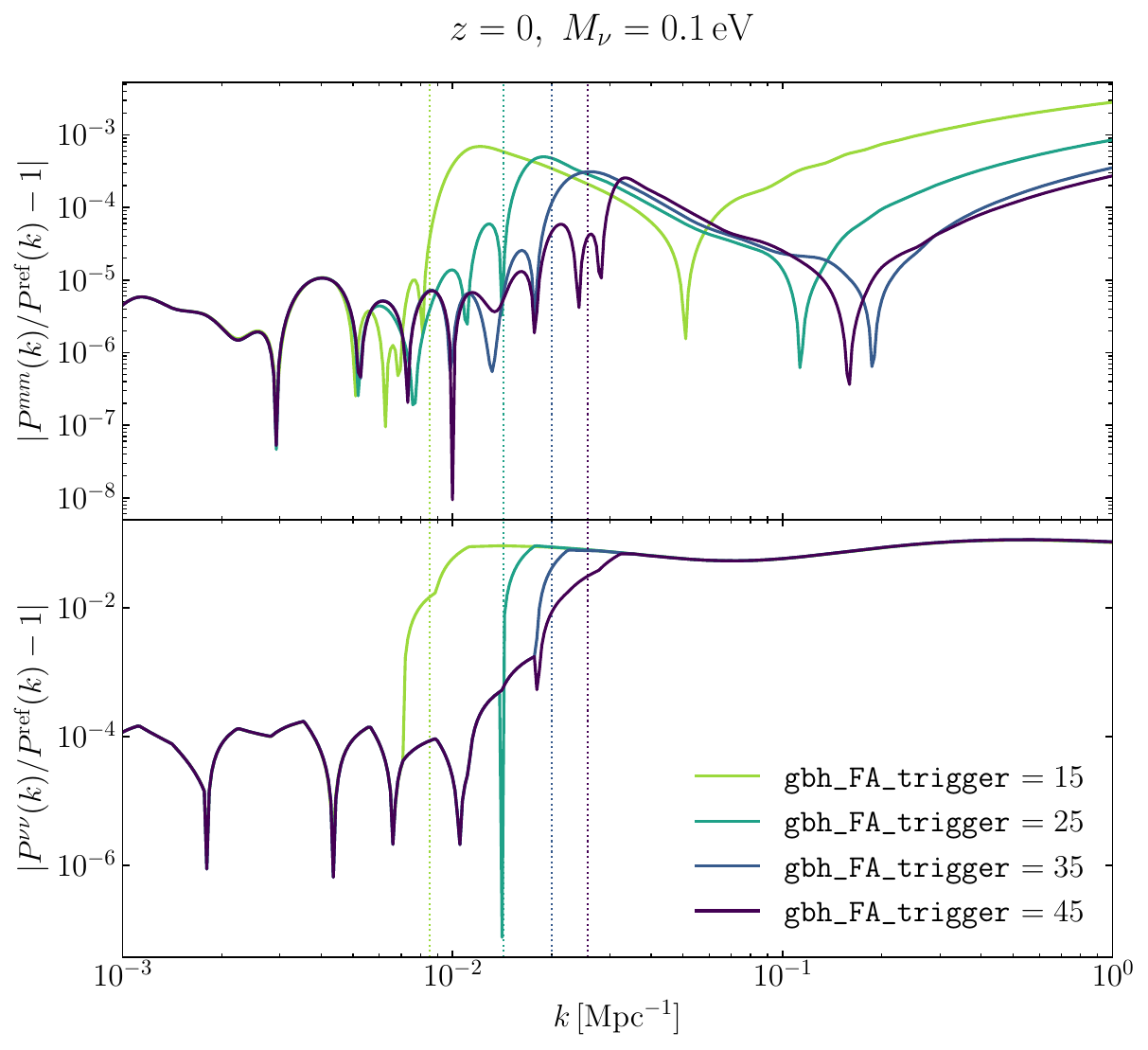}
    \caption{Top: Fractional error in the matter power spectrum compared to the reference \texttt{ncdm} run for various values of \texttt{gbh\_FA\_trigger}. At the default trigger 15, the error bumps up once at $k_{\rm switch}$. This is due to the neutrino fluid approximation taking over at $z=0$ for these modes. The error also keeps rising at very large $k$, which is due to the fluid approximation contaminating the gravitational potential at early times when neutrinos were relativistic, and then the resulting error propagating forward in time. Increasing the trigger pushes both of these bumps to smaller scales. Bottom: Fractional error in the neutrino power spectrum compared to the reference \texttt{ncdm} run for various values of \texttt{gbh\_FA\_trigger}. The error at small scales is dominated by the fluid approximation. Increasing the trigger moves this regime to smaller scales.}
    \label{fig:P_error_vs_trigger}
\end{figure}

\texttt{gbh\_FA\_trigger} controls when the switch from GBH to the DFA happens. By increasing it, every mode $k$ is integrated further in time using GBH before switching to the fluid approximation. For every \texttt{gbh\_FA\_trigger}, there is a mode called $k_{\rm switch}$ that would transition to the DFA exactly at the present time. For $k>k_{\rm switch}$, the transition happens before the present time, and for $k<k_{\rm switch}$, the transition never happens. Therefore, increasing \texttt{gbh\_FA\_trigger} increases $k_{\rm switch}$. This can be clearly seen in \cref{fig:P_error_vs_trigger}. Both the total matter and neutrino power spectrum maintain a high accuracy over a larger range of scales, before being dominated by the fluid approximation at $k>k_{\rm switch}$. 

Although massive neutrinos make up a small fraction of the present matter, increasing \texttt{gbh\_FA\_trigger} also has a considerable effect on the matter power spectrum according to the upper panel in \cref{fig:P_error_vs_trigger}. When neutrinos were still relativistic, they had a substantial effect on the gravitational potential. However, for large enough $k$, the code still transitions to the fluid approximation at these early times. The resulting inaccuracy in the neutrino overdensity is then imprinted on the gravitational potential and propagated forward in time to $z=0$. This was noticed in \cref{fig:gbh_spectra} as well, as a relatively higher error in the matter power spectrum compared to the \texttt{ncdm} code. Now, when we increase \texttt{gbh\_FA\_trigger}, these early-time inaccuracies are pushed to even smaller scales, and so the large $k$ matter power spectrum of our code becomes more accurate.

As briefly mentioned in \cref{sec:implementation}, the small-scale inaccuracies cannot be pushed to arbitrarily large $k$ simply by tuning up \texttt{gbh\_FA\_trigger}. Solving the GBH for longer demands larger $n_{\textrm{max}}$ and $\ell_{\textrm{max}}$ according to \cref{eqn:nmax_lmax_vs_y,eqn:conservative_trunc}. It was already hinted in \texttt{Paper I} that the GBH does not numerically converge for an arbitrarily large ODE system size. A complete mathematical understanding of this instability is yet to be figured out. Empirically, our numerical experiments show that as we increase \texttt{gbh\_FA\_trigger} beyond a certain mass-dependent threshold, first $k_{\rm switch}$ and then the adjacent modes of neutrino perturbations become unstable close to $z=0$. These are precisely the modes that are being integrated via GBH for the longest times. So a plausible explanation of this numerical instability would rely on the total evolution time or number of steps of the GBH system, consistent with the error introduced by the $n_{\textrm{max}}$ truncation accumulating over the integration [see the discussion below \cref{eq:trunc_nmax}]. Increasing \texttt{gbh\_FA\_trigger} further beyond the threshold either leads to very large errors in $P^{\nu\nu}$ at around $k\simeq k_{\rm switch}$ (for $M_\nu\lesssim 0.1\,\mathrm{eV}$), or leads to \texttt{NaN} values (for $M_\nu\gtrsim 0.1\,\mathrm{eV}$).

\begin{figure}[t]
    \centering
    \includegraphics[width=0.6\textwidth]{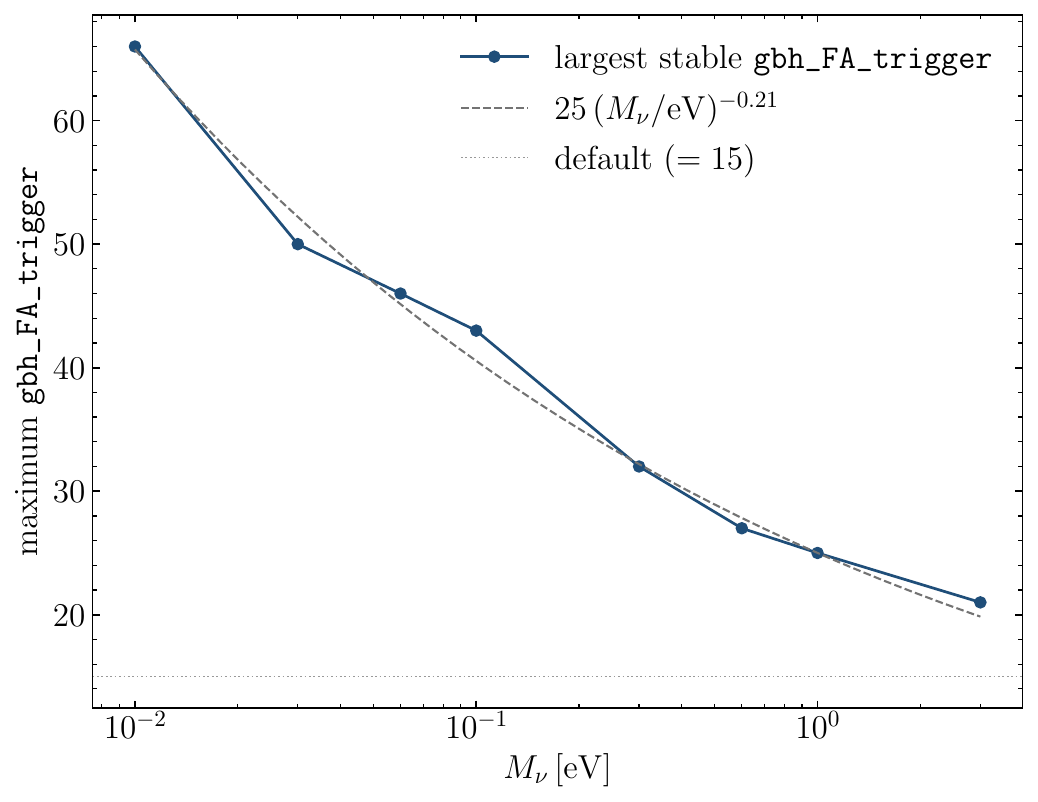}
    \caption{A conservative upper bound for \texttt{gbh\_FA\_trigger} vs neutrino mass. For choosing the blue dots for the given masses, we plot the error in neutrino power spectrum at $z=0$ close to $k_{\rm switch}$ against \texttt{gbh\_FA\_trigger}. At first, increasing \texttt{gbh\_FA\_trigger} stabilizes this error. However, as we increase the trigger further, it starts to deviate from its stable value, and ends up shooting up to large values or giving \texttt{NaN}. We select our upper bound conservatively at a \texttt{gbh\_FA\_trigger} where the error is just about to start to deviate from stability. The dashed curve here is a power-law fit to these empirical upper bounds.}
    \label{fig:trigger_ubound_vs_mass}
\end{figure}

So given this numerical instability, how far can we push \texttt{gbh\_FA\_trigger}? A conservative upper bound is given by
\begin{equation}
    \label{eqn:trigger_ub}\texttt{gbh\_FA\_trigger}\lesssim 25\left(M_{\nu}/\mathrm{eV}\right)^{-0.21}.
\end{equation}
This is found by demanding that the error in neutrino power spectrum close to $k_{\rm switch}$ remains stable under small changes in \texttt{gbh\_FA\_trigger}. The curve is depicted in \cref{fig:trigger_ubound_vs_mass}.

\begin{figure}[t]
    \centering
    \includegraphics[width=0.7\textwidth]{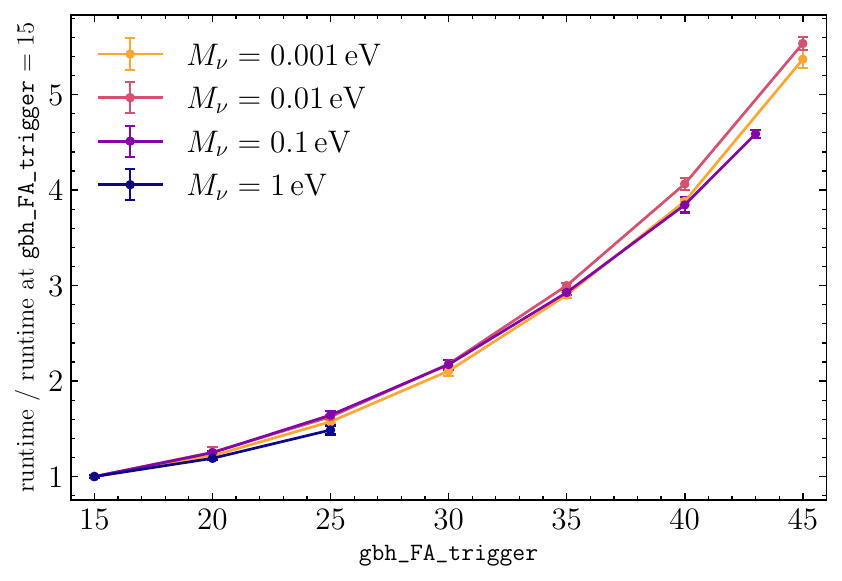}
    \caption{Runtime of GBH at different values of \texttt{gbh\_FA\_trigger} divided by its average runtime at \texttt{gbh\_FA\_trigger}$=15$. The error bars are one standard deviation found using 20 runs. Four different neutrino masses are plotted, and for each one, we have kept the trigger below the upper bound \eqref{eqn:trigger_ub}. The plot shows a steady rise in the runtime of the code, which is due to a larger trigger requiring larger $n_{\textrm{max}}$ and $\ell_{\textrm{max}}$ according to \cref{eqn:conservative_trunc}.}
    \label{fig:runtime_vs_trigger}
\end{figure}

Our default choice \texttt{gbh\_FA\_trigger}$=15$ is well below the above upper bound for all reasonable values of $M_\nu$. It gives an acceptable accuracy in all observables while minimizing the runtime of the code. As in \cref{fig:P_error_vs_trigger}, one can further decrease the error in matter power spectrum by increasing \texttt{gbh\_FA\_trigger} while keeping it below the upper bound~\eqref{eqn:trigger_ub}. However, remember that this increases the computational cost. \Cref{fig:runtime_vs_trigger} shows the runtime of the GBH code as a function of \texttt{gbh\_FA\_trigger}. The runtime rises steeply with the trigger, which is a result of the GBH system size growing quickly with the neutrino horizon, \cref{eqn:conservative_trunc}.

\section{MCMC validation against \ncdm{}}
\begin{figure}[t]
    \centering
    \includegraphics[width=\textwidth]{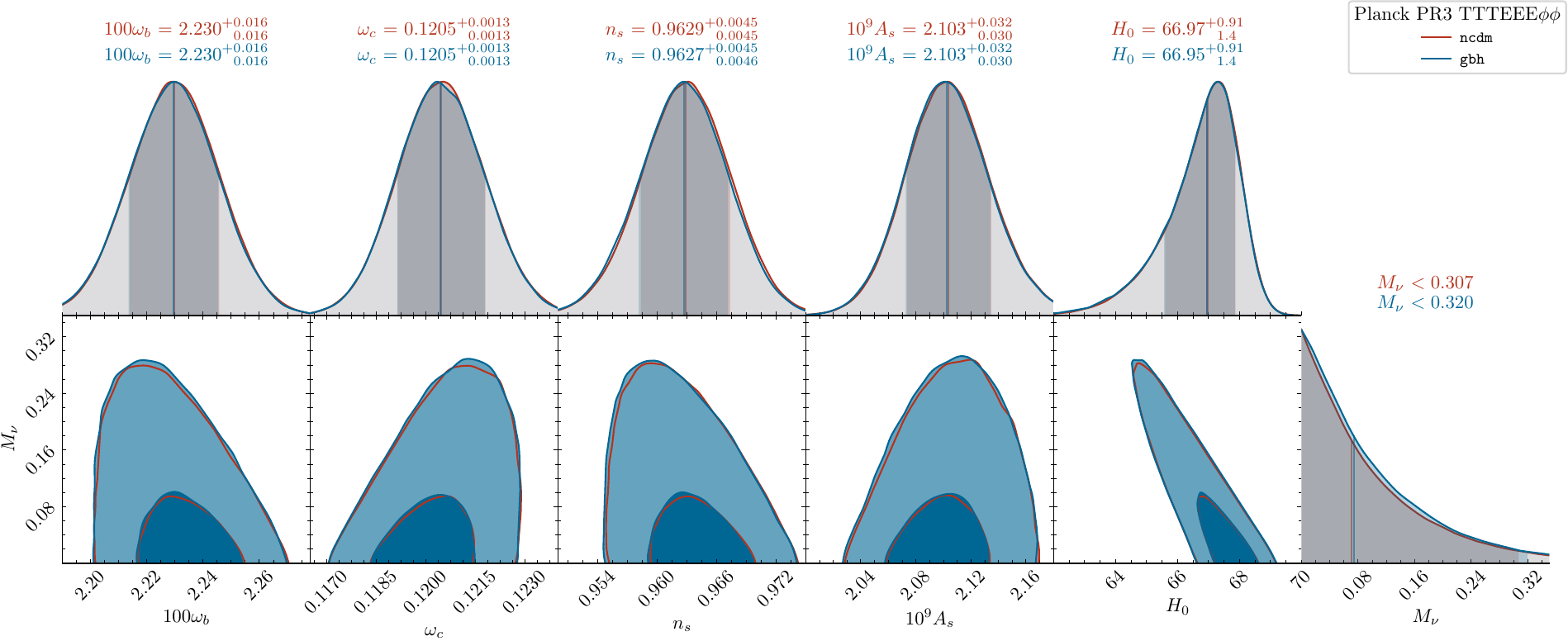}
    \caption{Marginalized posteriors from MCMC analyses of \Planck{} PR3 \texttt{TTTEEE}$+\phi\phi$ data, sampling $\Lambda$CDM$+M_{\nu}$, computed with the standard \class{} \ncdm{} module (red) and with the GBH$+$DFA implementation (blue). 
    Data, likelihoods, priors, and sampler settings are identical between the two, and only the treatment of the massive neutrino perturbations differs. 
    Top row: one-dimensional marginals for the $\Lambda$CDM parameters, with medians and $68\%$ quantiles annotated. 
    Bottom row: joint posteriors of $M_{\nu}$ with each parameter, and the $M_{\nu}$ marginal itself, with $95\%$ upper limits annotated. The contours indicate the 1 and 2$\sigma$ of the two-dimensional joint posterior density for pairs of parameters (i.e., the 39.3\% and 86.5\% mass levels).
    All median values agree to better than $0.05\sigma$ and the contours overlap throughout. }
    \label{fig:mcmc}
\end{figure}

\label[appendix]{app:mcmc}
The tests of \cref{sec:testing} establish that the GBH matches \ncdm{} at the level of transfer functions and power spectra. 
In this appendix, we validate that the GBH also reproduces the same cosmological parameter constraints as \ncdm{} when used in a full MCMC analysis. 
In \cref{fig:mcmc}, we show the results of two MCMC runs, one using \ncdm{} and the other using GBH$+$DFA, with identical data, likelihoods, priors and sampler settings (detailed below).
The two implementations are in excellent agreement, with marginalized medians agreeing to better than $0.05\sigma$ and contours overlapping throughout, including along the $M_{\nu}$--$H_{0}$ degeneracy. The $95\%$ upper limits on $M_{\nu}$ are also consistent between the two methods.

Our analysis uses \montepython{}~\cite{Audren:2012wb,Brinckmann:2018cvx} to sample the $\Lambda$CDM$+M_{\nu}$ parameter space, with the same priors and sampler settings for both \ncdm{} and GBH$+$DFA. We employ the \Planck{} PR3 temperature, polarization, and lensing nuisance-marginalized likelihoods~\cite{Planck:2018vyg, Planck:2019nip}, and sample the six $\Lambda$CDM parameters plus the sum of neutrino masses $M_{\nu}$.
Denoting a uniform distribution between $a$ and $b$ as $\mathcal{U}(a, b)$, we take flat priors for $\LambdaCDM$ parameters: $100\theta_s\sim\mathcal{U}(0.5, 10)$, $\omega_b\sim\mathcal{U}(0.005, 0.1)$, $\omega_c\sim\mathcal{U}(0.001, 0.99)$, $\ln \left(10^{10}A_s \right)\sim\mathcal{U}(1.61, 3.91)$, $n_s\sim\mathcal{U}(0.8, 1.2)$, and $\taureio\sim\mathcal{U}(0.01, 0.8)$.
We implement one massive neutrino species with a flat prior on the sum of neutrino masses $M_{\nu}\sim\mathcal{U}(0, 5)$ eV, fixing the other two species to be massless. 
We employ the Metropolis-Hastings algorithm in \montepython{} for sampling and a Gelman-Rubin convergence criterion of $R-1 \lesssim 0.01$, leading to effective sample sizes of at least 10,000.

\bibliography{gbh_paper.bib}

\end{document}